\documentclass{aa} 

\usepackage{natbib}
\usepackage{longtable}
\usepackage{lscape}
\usepackage{graphicx}
\usepackage{txfonts}
\usepackage[dvipsnames]{xcolor}
\usepackage{supertabular}
\usepackage{xfrac}  
\usepackage{multicol}
\usepackage[T1]{fontenc}

\bibpunct{(}{)}{;}{a}{}{,} 
\providecommand{\abs}[1]{\lvert#1\rvert}
\newcommand\tab[1][1cm]{\hspace*{#1}}
\newcommand\myfontsize{\fontsize{8.6pt}{8.6pt}\selectfont}

\begin{document} 

   \title{A search for Galactic post-asymptotic giant branch stars in \textit{Gaia} DR3\thanks{Tables A6 and A7 are only available in electronic form at the CDS via anonymous ftp to cdsarc.u-strasbg.fr (130.79.128.5) or via http://cdsweb.u-strasbg.fr/cgi-bin/qcat?J/A+A } }

   \titlerunning{Galactic post-AGB stars in \textit{Gaia} DR3}

  \author{I. Gonz\'alez-Santamar\'{\i}a\inst{1,2} \and M. Manteiga\inst{2,3} \and A. Manchado\inst{4,5,6} \and E. Villaver\inst{4,7} 
   \and  A. Ulla\inst{8,9} \and C. Dafonte\inst{1,2}}
  % \authorrunning{I. Gonz\'alez-Santamar\'{\i}a et al.}
 \institute{Universidade da Coru\~na (UDC), Department of Computer Science and Information Technologies, Campus Elvi\~na s/n, 15071 A Coru\~na, Spain \\
  \email{iker.gonzalez@udc.es}
  \and
  CIGUS CITIC, Centre for Information and Communications Technologies Research, Universidade da Coru\~na, Campus de Elvi\~na s/n, 15071 A Coru\~na, Spain                    
  \and
   Universidade da Coru\~na (UDC), Department of Nautical Sciences and Marine Engineering, Paseo de Ronda 51, 15011, A Coru\~na, Spain \\
   \email{manteiga@udc.es}
  \and
  Instituto de Astrofísica de Canarias, 38200 La Laguna, Tenerife, Spain
  \and
  Universidad de La Laguna (ULL), Astrophysics Department, 38206 La Laguna, Tenerife, Spain
  \and 
  CSIC, Spain
  \and 
 Agencia Espacial Espa\~nola, 41015 Sevilla, Spain
 \and
  Universidade de Vigo (UVIGO), Applied Physics Department, Campus Lagoas-Marcosende, s/n, 36310 Vigo, Spain
  \and
  Centro de Investigación Mari\~{n}a, Universidade de Vigo, GEOMA, Edificio Olimpia Valencia, Campus Lagoas-Marcosende, 36310 Vigo, Spain
  \\
           }
   \subtitle{}

   \date{Received 6 March 2023 / Accepted 9 June 2024}
\abstract
  % context heading (optional)
   {When low- and intermediate-mass stars leave the asymptotic giant branch (AGB) phase, and before they reach the planetary nebula stage, they enter a very brief and rather puzzling stellar evolutionary stage called post-AGB stage. The post-AGB phase 
   lasts very briefly, about a few 
   thousand years at most. The number of objects that are confirmed in this phase therefore is really small, and our understanding of this elusive stellar evolutionary stage is accordingly very limited.
   }
  % aims heading (mandatory)
   {We provide a reliable catalogue of Galactic post-AGB stars together with their physical and evolutionary properties obtained through \textit{Gaia} DR3 astrometry and photometry. As an added product, we provide information for a sample of other types of stellar objects, 
   whose observational properties mimic those of post-AGB stars. 
   } 
  % methods heading (mandatory)
   {Post-AGB stars are characterised by their infrared excesses and high luminosities. The publication of precise parallaxes in \textit{Gaia} DR3 made it possible to calculate accurate distances and to revise the derivation of luminosities for post-AGB candidates, so that objects outside the expected luminosity range can be discarded. We started by identifying post-AGB stars or possible candidates from the bibliography, and we then searched for their \textit{Gaia} DR3 counterpart sources. Using the available photometry, interstellar extinction, spectroscopically derived temperatures or spectral types and parallax-derived distances from the literature, we fitted their spectral energy distributions and estimated their luminosities and circumstellar extinctions. By a comparison to models, the luminosity values allowed us to determine which objects are likely post-AGB stars from other target types.
   Their position in the Hertzsprung-Russell diagram allows a direct comparison with updated post-AGB evolutionary tracks and an estimation of their masses and evolutionary ages.}
  % results heading (mandatory)
   {We obtained a sample of  69 reliable post-AGB candidates that meet our classification criteria, which provide their coordinates, distances, effective temperature, interstellar and circumstellar extinction, luminosity, mass, and evolutionary age. In addition, we provide similar data for other stellar objects in our initial compilation,
   such as supergiant stars and young stellar objects. Our identifications and parameters are compared with others found in the recent literature for the subject.}
  % conclusions heading (optional)
   {We selected the data with the best precision in parallax and distance to obtain more accurate luminosities, which allowed us to confidently classify the objects of the sample in different stellar phases. In turn, this allowed us to provide a small but reliable sample of post-AGB objects. The derived mean evolutionary time and average mass values agree with theoretical expectations and with the mean mass value obtained in a previous work for the subsequent evolutionary stage, the planetary nebula stage. % that hopefully will help to improve our understanding of this brief and poorly known stage in the late evolution of stars.
   }

        
\maketitle

\section{Introduction}

The stellar phase known as post-asymptotic giant branch (post-AGB) stage is a very fast (a few thousand years) and quite unknown phase that takes place at the end of the lifetimes of low- and intermediate-mass stars, after the AGB phase and before the planetary nebula stage, in which the star ionises the previously ejected envelope. The beginning of the post-AGB phase is not exactyl determined, and its onset depends on the stellar mass and metal content. Furthermore, the departure from the AGB phase is defined somewhat arbitrarily in the stellar evolution models. For \cite{1993ApJ...413..641V} the AGB mass loss terminates when the envelope mass is reduced to a value at which the stellar effective temperature increases beyond a reference value by an amount of $\Delta$ log(Teff)=0.3. Other authors have different criteria. \citet{2016A&A...588A..25M} set the onset at the point in time when as a result of stellar winds, the H-rich envelope mass drops below 1\% of the stellar mass. 

During the post-AGB phase, the star evolves at an almost constant luminosity towards hotter effective temperatures, while its envelope expands into the interstellar medium. Due to the high stellar temperature, this envelope composed of gas and dust begins to be ionised (see e.g. \citealt{2002ApJ...581.1204V}). At this moment, the star enters the protoplanetary nebula phase. As in \cite{2021A&A...656A..51G}, we can
assume a minimum stellar temperature of 13 000 K for a transition stage from a preplanetary nebula (Weidmann et al. 2020) to about 24 000 K for a complete ionisation of the nebula (Kwok, 2000).

Several studies have recently been carried out to identify post-AGB stars with the aim to better characterise this brief stellar phase. \citet{1997A&AS..126..479G} identified 110 possible post-AGBs based on their IRAS\footnote{InfraRed Astronomical Satellite} infrared colours, most of which lack an optical counterparts. At the beginning of this century, \citet{2006A&A...458..173S} provided optical counterparts for more than 100 post-AGB candidates that were previously proposed as such based on their IRAS fluxes. Almost in parallel, a more extensive catalogue of post-AGB candidates was presented by \citet{2007A&A...469..799S}, which was later extended in \citet{2012IAUS..283..506S}. This is known as the Torun catalogue of post-AGB stars and is the largest catalgoue to date. It contains 296 sources classified as either likely or possible post-AGBs. 
The sample of post-AGB stars has been expanded to the Large Magellanic Cloud (LMC) by \citet{2015MNRAS.454.1468K}. They also included post-red giant branch (RGB) stars. These stars are thought to be produced from binary interactions in the RGB phase, and their observational properties mimic those of post-AGB stars,  although they are not expected to reach luminosities as high as those of post-AGB stars \citep{2015MNRAS.454.1468K}. In a previous article \citep{2021A&A...656A..51G}, we found that a fraction close to 50\% of the central stars of PNe are red and might therefore be unresolved binary systems. Finding the relative number of stars of one type and of
%individual 
single or binary type at these evolutionary stages can give us clues about the fraction of binary stars that has evolved companions.

 \citet{2022ApJ...927L..13K} recently studied the evolutionary state of 31 Galactic post-AGB candidates with chemical abundance information. By using \textit{Gaia} EDR3 distances and known photometry, they built the spectral energy distribution (SED) for these objects, and by fitting them to models, they estimated their luminosities and temperatures. It is important to note that 20 of the objects they analysed have quite poor astrometry in \textit{Gaia} DR3. 
%, by using Gaia Early Data Release 3 (EDR3). 
%From their sample, 20 objects have very poor astrometry in DR3 and 11 objects are in common with our sample. In the present work, we have classified seven out of these 11 objects as post-AGB, three objects as post-RGB and one as supergiant star. 

%Both the temperatures and luminosities obtained in \citet{2022ApJ...927L..13K} are in general higher than those obtained by us.

%{\bf EV: Su muestra final es similar a la tuya en tamaño, ellos tienen unos 30 objetos y además hacen análisis de la SED. Tienes objetos en comun? que puedes decir de ellos?}
% Iker: Ya lo decimos en la seccion 3, cuando analizamos los resultados. Tambien es necesario ponerlo aqui?
\citet{2020PASJ...72...99P} analysed the properties of 8 post-AGB candidates based on \textit{Gaia} DR2 astrometric and photometric data. Also using \textit{Gaia}, in this case, DR3, \citet{2022MNRAS.516L..61O} investigated their nature as possible post-AGB of 249 objects. It is noteworthy that most of the objects they selected have large uncertainties in their parallaxes, which leads to very unreliable distance values. They considered good parallaxes to have a relative error ($1\sigma$) between 10\% and 100\%. Additionally, we detected some inconsistencies in their study on which we comment in section 3. % and compare in detail our sample with the one in \citet{2020PASJ...72...99P} and consequently, a high uncertainty in their parallax measurement. %located 134 galactic post-AGB candidates in a Hertzsprung-Russel (HR) Diagram, discovering that 53 of them are actually post-RGB stars.

Finally, in a recent work, \citet{2022PASJ...74.1368A} studied the evolutionary state of 20 post-AGB candidates by using \textit{Gaia} DR2 and EDR3.
%In section 3, we discuss our identification of post-AGB phase objects versus that carried out in these works, as well as the physical properties derived for the objects.

We aim to identify and analyse the properties of bona fide post-AGB stars in more detail by selecting objects with accurate astrometric measurements in \textit{Gaia} DR3, which allows us to precisely locate them in the Hertzsprung-Russell (HR) diagram. This consequently leads to a quite reliable classification  of these stars as post-AGB objects. If we were to select objects from Galactic post-AGB stars candidates with good astrometry in \textit{Gaia} DR3, we would exclude in a first approximation most astrometric binaries. The general rule is that the threshold of the \textit{Gaia} astrometric quality parameter called renormalised unit weight error (RUWE) $\leq$ 1.4 is used to indicate single well-behaved solutions (\citealt{2018A&A...616A...2L,2021A&A...649A...4L}). The inconsistency of source observations with the \textit{Gaia} astrometric five-parameter model could be caused by binarity (\citealt{2018A&A...616A...2L}) or other factors that cause the photocentre of the source to wobble during the \textit{Gaia} observation window. In summary, a restrictive criterion in the astrometric quality helps us to select objects that are more likely to be individual sources, and on the other hand, it helps us to estimate their luminosity and evolutionary stage better. The incidence of binaries in our resulting sample is addressed by studying the SED morphology, and we also consider this in the context of the literature.

We started by collecting a sample of post-AGB candidates that was as complete as possible from the currently available catalogues (Sect. 2), and we implemented restrictive filtering over the astrometric quality of the objects so that we only kept those with the most accurate distance values. We gathered a general sample of 964 post-AGB candidates from the literature, out of which we filtered a subset of 178 objects with accurate astrometric measurements in \textit{Gaia} DR3. A good distance determination 
is not enough to obtain a reliable adjustment of the SED; additional information about the temperature and/or extinction in the direction of the source is required for a consistently derived luminosity of the object. To better constrain the value of the total extinction, interstellar and circumstellar, in the direction of the source, we opted to limit our work to objects with available interstellar extinction measurements in the literature. Of the previous sample of 178 objects, we kept 146 Galactic post-AGB candidate stars with literature values of their interstellar extinction. In this last sample, the information available in the Simbad database as well as images of every sky field in the Aladin Sky Atlas \citep{2000A&AS..143...33B} were analysed. We finally discarded 28 sources from the further analysis because they were either already classified as PNe (5 objects), had higher effective temperatures in the literature than  24 000 K (11 objects), or because the identification of the optical counterpart was dubious (12 objects). Our final working set 
%finally 
was 118 objects.
 We found that the temperatures for about 67\% of them were derived from spectral analysis, 
 %and 
and some of these from a high-resolution spectral analysis. The average temperatures for the remainder corresponding to their spectral types were used. Section 3.2 describes the problems associated with the different quality of the temperature determinations we used in detail. %Also, a well-studied dwarf nova was excluded. 

Photometry in a wide spectral region compiled by the Spanish Virtual Observatory SED Analyser (VOSA\footnote{http://svo2.cab.inta-csic.es/theory/vosa/}) allowed us to build the 
SED for each object. The knowledge of distances and temperatures allowed us to obtain luminosities and to estimate the total extinction from the SED fitting (Sect. 3.3). We used the current knowledge of interstellar extinction in the direction of every object to verify the consistency of the total extinction we obtained in the fits.

Based on evolutionary tracks, we used the distribution of objects in an HR diagram to confirm 69 objects out of 118 candidates as post-AGB stars. Some other objects could be classified as horizontal branch stars (3 objects), luminous supergiants (3 objects), young stellar objects (YSOs; 14 candidates), and the remaining objects are unconfirmed post-AGB candidates. The details are given in Sec. 3.5. 
%\textcolor{red}{(Arturo/Ana/Eva ?) Esto no se entinde. Quieres decir que otros son possible post-AGB? Yo lo escribiria; being some possible pots-AGBs,supergiants, post-RGB and unclassified stars } 
In Section 3.6, this classification is compared with the classifications presented in other recent papers about the subject. %\textcolor{red} {The evolutionary nature of the remaining stars could not be confirmed.} 
In Sect. 4, we focus on analysing the evolutionary properties of the set of 69 objects that we identify as belonging to the post-AGB stage. In Sect. 5, we comment on some interesting objects, and in Sect. 6, we summarise our conclusions.

\section{Sample selection: Method}
The first step in this research was to collect all the objects from the literature that were catalogued as confirmed or possible post-AGB stars. For this purpose, we used the online\footnote{https://fox.ncac.torun.pl/camkweb/postagb2.php} Torun catalogue of %galactic 
Galactic post-AGB stars (\citealt{2012IAUS..283..506S}), the Simbad astronomical database, and the spectroscopic atlas of post-AGB and planetary nebulae by \citet{2006A&A...458..173S} (from now, the on Suarez et al. catalogue). 

We gathered all objects that were catalogued as likely (209) or possible (87) in the Torun catalogue, as post-AGB (331) or post-AGB candidate (507) in the Simbad database, and as post-AGB (102) in the Suarez et al. catalogue. To combine the three catalogues, we first matched the Torun objects with those in Simbad using a cross-match radius of one arcsecond. As a result, we obtained a set of 929 objects. Subsequently, we matched them with Suarez et al. catalogue (by the same method and using coordinates from Torun when 
%it was 
possible), obtaining a final sample of 964 objects.

The next step was to cross-match our list of objects with the \textit{Gaia} DR3 archive to obtain their parallaxes and distances. %Once we had this general sample of post-AGB stars (or candidates), the idea was to identify them as Gaia DR3 sources, in order to obtain their parallaxes, and consequently their distances. The knowledge of precise distances to stars is very useful, as it allows us to estimate their intrinsic properties from the observational ones. With the aim of obtaining reliable Gaia source identifications for all the stars in the sample, we decided to 
We again used a search radius of 1 arcsecond from the literature coordinates. In some cases, the coordinates from Simbad or from the Suárez et al. catalogues differ slightly from the coordinates from the Torun catalogue. This can lead to discrepancies in the identification of the \textit{Gaia} DR3 source. We decided to prioritise the Torun coordinates because they come from a revised compilation of post-AGB stars. As a result, we were able to identify 843 objects as \textit{Gaia} DR3 sources. This is about 87\% of the whole sample. %Then, as it is necessary to know the parallax of a source to estimate its distance,  
Finally, we discarded objects with unknown distances in \citet{2021AJ....161..147B}, which left 765 objects. 

In \textit{Gaia} DR3, the parallaxes ($\pi$) show a bias or zero point ($z_{0}$) that should be considered and subtracted from the measured value to obtain the actual parallax ($\pi_{0}$). According to \citet{2021A&A...649A...4L}, this zero point has a mean value of -17 $\mu$as, although its value varies depending on the celestial position, colour, and magnitude of the star. The \textit{Gaia} web provides a Python code for estimating this zero point \footnote{https://gitlab.com/icc-ub/public/gaiadr3\_zeropoint}. After obtaining the zero points, we corrected the parallaxes with the following simple expression:

$$\pi_{0} = \pi - z_{0}.$$

The uncertainties in the \textit{Gaia} parallaxes include the value given in the \textit{Gaia} DR3 archive, the internal uncertainty, and a systematic uncertainty that depends on the source brightness. We followed the prescriptions given in \citet{2021A&A...649A...5F} to estimate the total parallax uncertainties of our objects.

From the inverse of the parallaxes, it is possible to estimate the stellar distances, but this simple approach is only valid for objects with relative uncertainties as low as 10\%. In general, this is not the case for our data, and we therefore used the distances calculated by a Bayesian approach for Milky Way stars by \citet{2021AJ....161..147B}, which consists of assuming an a priori probability volume density of stars in the Galaxy that decreases exponentially on an appropriate distance scale. This method not only provides an estimated distance for a source, but also gives low and high error distance bounds. %So with these parameters we also calculated the distance low and high uncertainties.

%In order to analyse in detail the properties of these stars, 
To infer useful stellar properties that depend on distance, such as luminosity, it is important to have precise distances to these objects. We therefore decided to apply filtering to our sample according to the distance uncertainties and the astrometric quality. Astrometric quality indices such as the unit weight error (UWE; see the official \textit{Gaia} website) and RUWE beyond certain boundaries prevent \textit{Gaia} users from errors in the astrometric solution that can be due to binarity or to irregularities in the fitted source. We decided to use the same filtering criteria as we used in one of our previous works, where we conducted a similar study, but on the Galactic sample of central stars in planetary nebulae  (\citealt{2021A&A...656A..51G}). This filtering consisted of the following constraints: The relative error in parallax and distance (for the lower and upper bounds) below 30\% and the astrometric quality parameters UWE and RUWE below certain threshold values that are recommended in the \textit{Gaia} documentation (UWE<1.96 or RUWE<1.4). After applying these constraints, the set contained 178 objects that we consider to have good astrometric measurements in \textit{Gaia} DR3.

% Quitar las de las Nubes de Magallanes

\begin{figure}[h!]
        \centering
        \includegraphics[width=8.5cm,height=6cm]{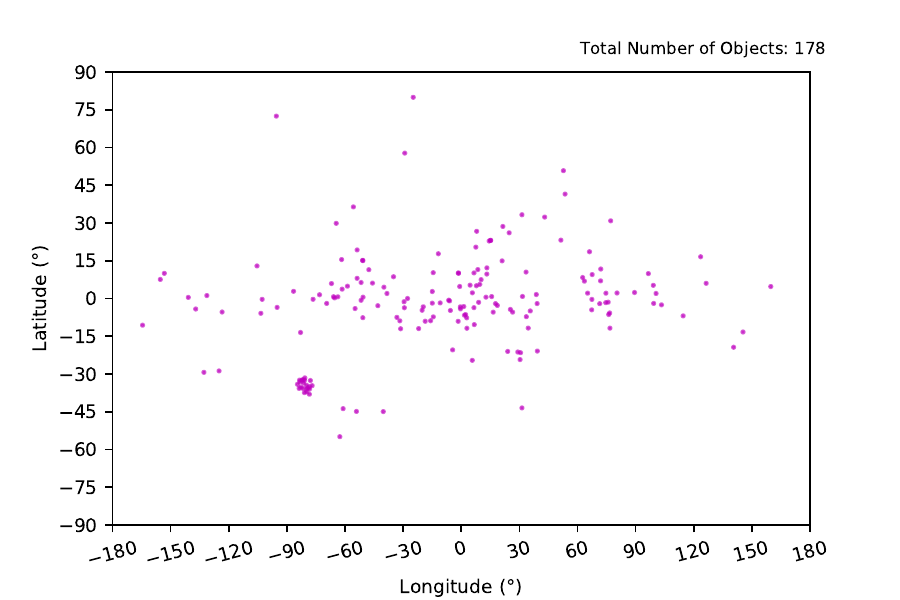}
        \caption{Galactic distribution of the 178 post-AGB candidates from the selected sample.}
        \label{fig:galactic_distribution}
\end{figure}

The Galactic distribution of these stars %(in longitude and latitude coordinates), 
shows a small cluster of 21 objects in a region of mid-south latitudes (see Fig. \ref{fig:galactic_distribution}) at longitudes coincident with those of the direction of the LMC (Large Magellanic Cloud). 
Furthermore, we verified that all of these objects were included in the \citet{2011yCat..35300090V} catalogue of LMC post-AGB stars, and 10 of them are also included in \citet{2015MNRAS.454.1468K}.
We therefore decided to exclude these stars from our present study and focused %The stars contained in that region could be integrated in the LMC or only in the line of sight of it.
%, but it is not easy to discern the difference.
% Explicar proceso de diferenciacion de las que estan en LMC
%As the stars from LMC could follow another evolutionary rules, 
%we decided to separate provisionally these 20 stars from our sample, and we focused 
on the remaining 157 Galactic post-AGB candidates.

In Table \ref{table:candidates} we provide general data of the objects in this latter sample: a running number, the name, the \textit{Gaia} DR3 ID, the J2000 coordinates, the G magnitude, the interstellar extinction in the V band, the spectral type, the reference for an identification as post-AGB (the Torun, Simbad, and Suárez et al. catalogues), and a flag providing information about binarity, variability, and the reason for excluding the source from the further analysis, as described in the previous section, together with a reference.

\begin{figure}[h!]
        \centering
        \includegraphics[width=8.5cm,height=6cm]{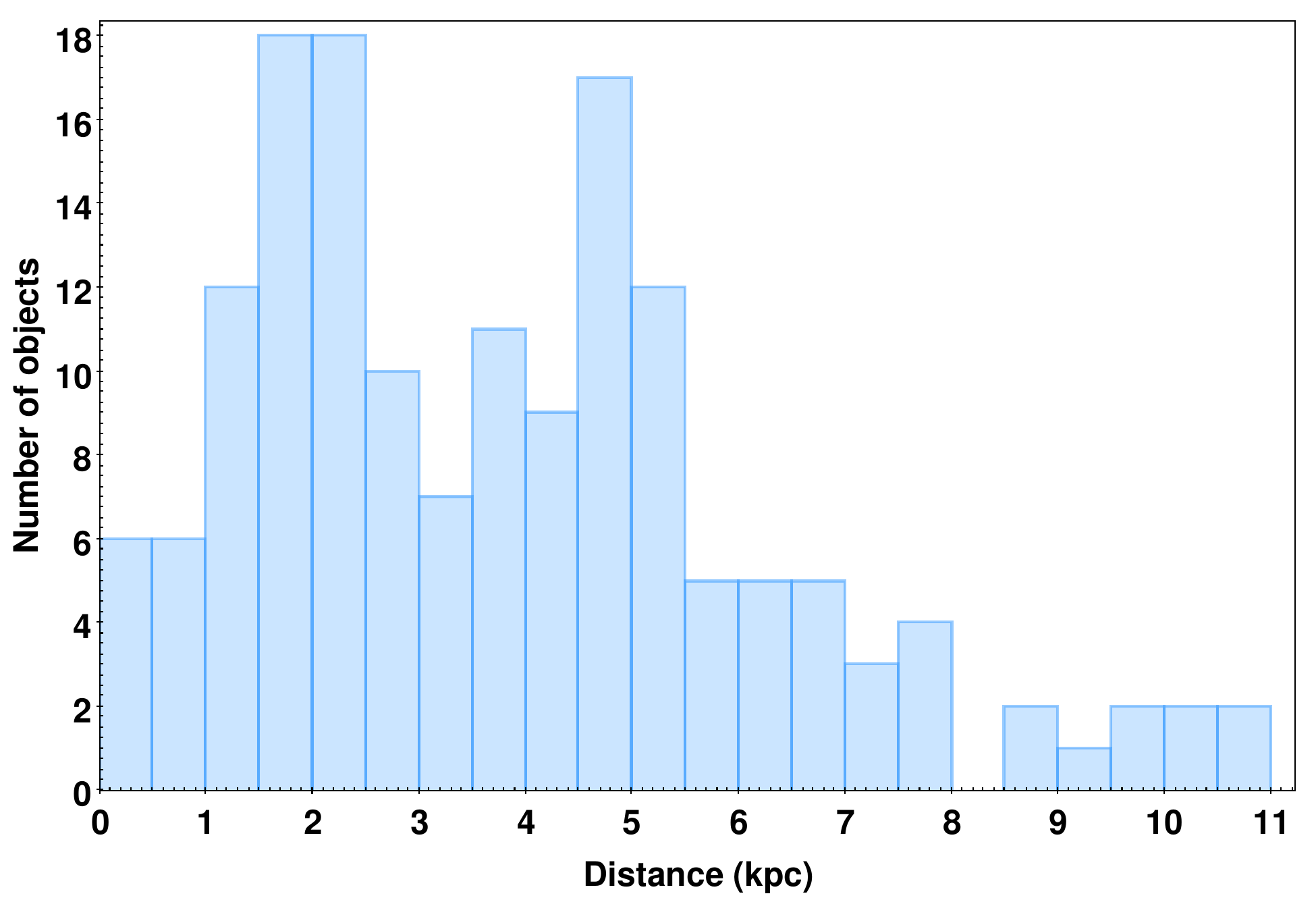}
        \caption{Distance to the 157 Galactic post-AGB candidates with good-quality astrometry.}
        \label{fig:distance}
\end{figure}

%To investigate the completness of our sample, we analyse their distance distribution 
The distance distribution of all these post-AGB candidates is shown in Fig. \ref{fig:distance}. We did not attempt to analyse the completeness of the sample since it presents several observational biases, including the impossibility of detecting the optical counterpart of some of the infrared sources with data in the IRAS catalogue.%While most of the stars are located closer than 5 kpc, there are several objects reaching a distance beyond 10 kpc. The distribution presents two picks, one around 2 kpc and the other around 5 kpc, which make it  difficult to estimate the completeness of the sample.  {\bf en realidad no dices nada de completitud....quedate en un bin cercano y compara con la poblacion que esperas de post-agbs comparado con la de enanas blancas o algo asi....pero tal y como está no dices nada}

Individual distance values in pc, except for discarded objects, are listed in the second column of Table \ref{table:postAGB}, Table \ref{table:postRGB}, Table \ref{table:YSO} and Table \ref{table:supergiant}.

%\section{Identifying Galactic post-AGB stars}
\section{Identification of Galactic post-asymptotic giant branch stars}
\begin{figure}[h!]
        \centering
        \includegraphics[width=8.5cm,height=6cm]{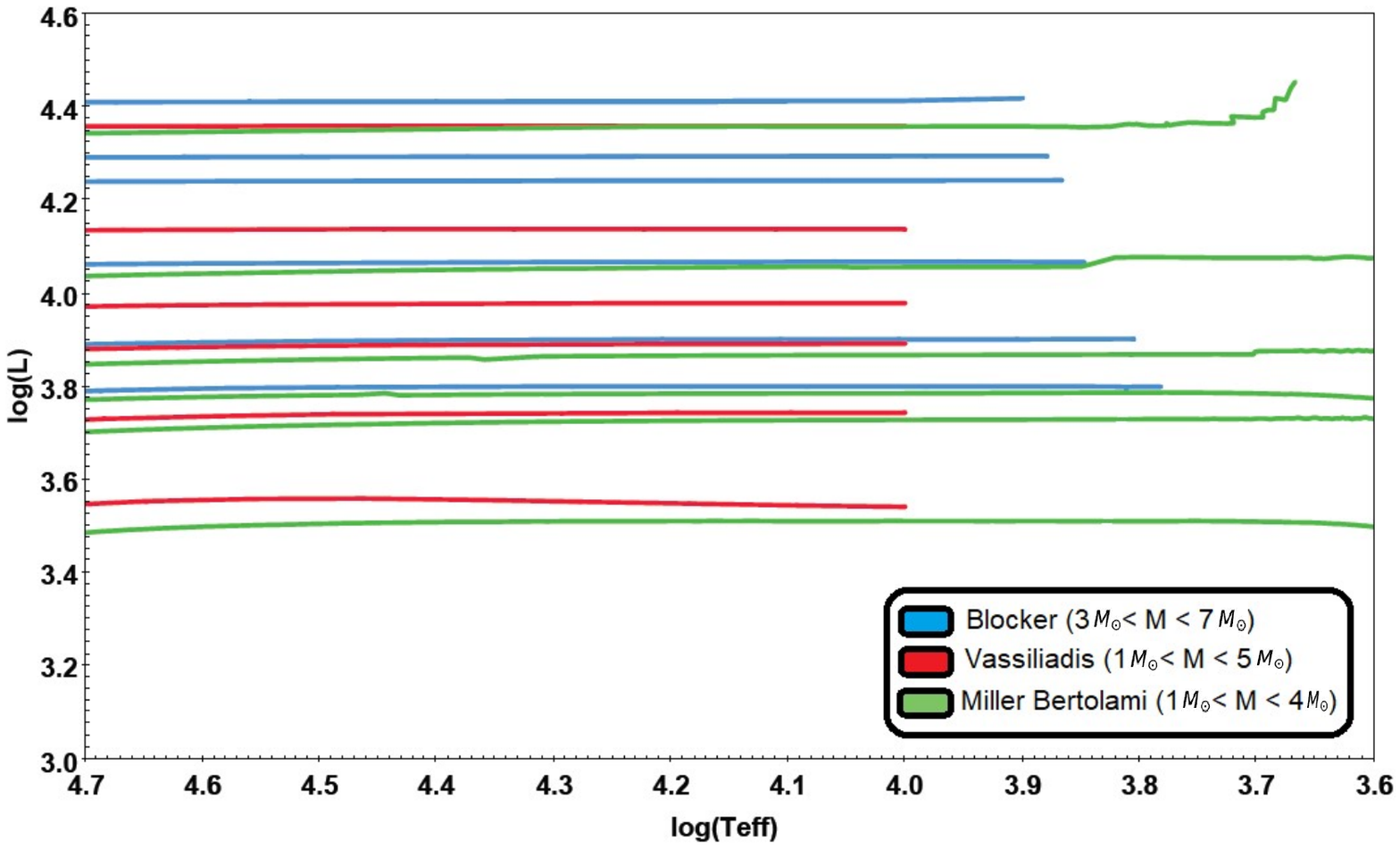}
        \caption{Region in the HR diagram covered by different evolutionary tracks of post-AGB evolution.}
        \label{fig:tracks}
\end{figure}

The luminosity of a star is a very useful property for distinguishing between post-AGB stars and other stellar objects with similar colours that are located beyond the main sequence (MS) of the HR diagram, as is the case for YSOs. 
To narrow the luminosity range that corresponds to post-AGB stars, we resorted to different evolutionary models for hydrogen-burning post-AGB stars: The classical models by \citet{1993ApJ...413..641V} (for masses between 1 and 5 $M_{\odot}$) and by \citet{1995A&A...299..755B} (for masses between 1 and 7 $M_{\odot}$), and more recently, the model by \citet{2016A&A...588A..25M}, which includes different metallicities (for masses between 0.8 and 4 $M_{\odot}$). Although the range of masses of the progenitor stars is different in each of the models, it should be noted that the evolution of stars with masses greater than 4 $M_{\odot}$ is very fast. 
%it is expected that the statistics will be very poor 
Rather poor statistics is therefore expected 
for objects of these and higher masses. Their expected number is probably also very small given the initial mass function. The models (see Figure \ref{fig:tracks}) agree reasonably well on a luminosity range of 
%between  
%$\log(\frac{L}{L_{\odot}})=3.4$ and $\log(\frac{L}{L_{\odot}})=4.5$,
$3.4\leq \log(\frac{L}{L_{\odot}})\leq 4.5$. 

This luminosity range for post-AGB stars meets the criteria used by \citet{2015MNRAS.454.1468K}, 
%that considers 
who considered that the luminosity range for post-AGB stars is between 2500 $L_{\odot}$ and 35,000 $L_{\odot}$. 
%( $3.4 < \log(\frac{L}{L_{\odot}}) < 4.5$). 
According to these authors, objects above this upper limit may be supergiants or hypergiants, high-mass stars that quickly initiate helium-core fusion after they have exhausted their hydrogen and that continue to fuse heavier elements after helium exhaustion until they develop an iron core, at which point, the core collapses to produce a Type II supernova. In contrast, objects below the lower limit may be post-RGB stars, YSOs, or they may be other evolved stars such as horizontal branch (HB) stars. %Among this last group of objects, it is expected that all those below 100 $L_{\odot}$ ($\log(\frac{L}{L_{\odot}})<2$) are concretely YSOs. 

The main problem in classifying high-luminosity stars as either post-AGB evolved objects or high-luminosity massive objects is that the two types of objects share many observational features: the optical spectra are similar, they have unstable and extended atmospheres, their gas-dust envelopes expand, they have a high IR excesses, and their IRAS colours are similar \citep{1997A&AS..126..479G}. This matter is dealt with in detail by \citet{2018ARep...62...19K}, who argued the need to determine and compare various parameters: the position in the Galaxy, the luminosity, the wind parameters, the SED, and the chemical composition to allow for an accurate classification. For this reason, the few objects that we found up to the limit $\log(\frac{L}{L_{\odot}})=4.5$ are discussed individually (see Sect. 5).

Objects below the minimum luminosity indicated by post-AGB evolution models present a different problem. These objects have  $2 < \log(\frac{L}{L_{\odot}}) < 3.4$. As we mentioned before, in \citet{2015MNRAS.454.1468K}, the post-AGB candidates in the Magellanic Clouds located in this 
%region 
range were tentatively identified as post-RGB stars. According to these authors, these stars are most likely the result of binary interaction in which their evolution towards the AGB is interrupted, but only in some cases was their binary nature confirmed, and some of them might equally be  the result of a merging process. It is beyond the 
%objective 
scope of this paper to analyse these objects,
and for this reason, we refer to them as unconfirmed post-AGB candidates. %post-RGB objects in \citet{2015MNRAS.454.1468K} are located 
%We will follow this guideline to confirm the identification of post-AGB candidates within our sample.  

Thus, to classify our objects as post-AGB stars or as another type of stellar objects, it is necessary to calculate accurate luminosities and also to have reliable estimates of their temperatures. 

\subsection{Interstellar reddening}
To explore the best possible determination of the interstellar extinction values for our objects, we used \textit{Gaia} DR3 coordinates and distances and searched the bibliography for the corresponding extinction values. After analysing and comparing data from different catalogues and dust maps, we decided to use the extinction values from \citet{2019yCat.4038....0S}. We obtained $E(B-V)$ values from this catalogue by performing a 
%X-match 
cross-match between our objects sample and the TESS\footnote{Transiting Exoplanet Survey Satellite} Input Catalog v8.0 (\citealt{2019yCat.4038....0S}) using the Topcat tool (\citealt{2005ASPC..347...29T}). The Stassun catalogue contains extinctions from dust maps for 146 objects of our sample. %, that is 93\% of it. 
We always used distance-dependent extinction values when they were available. Extinction errors are provided for 116 objects, and the magnitudes of about 80\% of them are below $\Delta A_{V}=0.15$. The interstellar extinction values are listed in Table \ref{table:candidates}.

\subsection{Effective temperatures}
We searched the Simbad database for temperature values and references for each object in our sample of 146 candidates. As a result, we discarded 28 objects that were too hot or for which the identification of the central source in the literature was uncertain. In consequence, we ended up with a sample of 118 objects, that we considered our final sample. The details are included in  Table \ref{table:candidates}. We found very different temperature values. Approximately 67\% of objects have precise temperature determinations that come from spectral analysis. This is the case of 11 objects in common with the work of \citet{2022ApJ...927L..13K}, or those in common with \citet{2023A&A...674A.151C} or \citet{2012A&A...543A..11M} (see references in Tables A.2, A.3, A.4 and A.5).  

For several cases, the T effs come from spectral types derived from medium-resolution spectra, as in the \citet{2006A&A...458..173S} catalogue.
For other cases, only Simbad spectral types are available, some of which cover a quite wide range of subtypes or even types. We also note that 
for some objects, the spectral classifications in the MK system, which are generally old, are quite discrepant with the spectroscopic temperatures obtained in more recent publications. For instance, the star BD+48 1220 is assigned spectral type A4Ia (8550 K) in Simbad based on \citet{1965LS....C05....0H}, while \citet{2019ApJ...879...69T} reported a value of 6389 K from an APOGEE spectra analysis.  This leads us to deduce that at least for some cases, the effective temperatures obtained from spectral types may be inaccurate.

Following the precision of the literature values, the temperatures from spectral analysis were prioritised over temperatures obtained by spectral classification in MK types, which in turn were prioritised over average temperatures obtained directly from the spectral type in the Simbad database.
Tables A.2, A.3, A.4, and A.5 list the effective temperature we adopted for each object together with a reference and a flag indicating its origin. 

%\textcolor{red}{In section 1, we mentioned that 28 objects out of the sample of 146 post-AGB candidates, were excluded for further analysis for different reasons, so we ended up with a sample of 118 objects, that we will consider our final sample.}

\subsection{Luminosity and total extinction from fitting the spectral energy distribution}

To estimate the luminosity of a star, its bolometric flux or magnitude, distance, and interstellar extinction values are needed. A simple approach consists of obtaining the stellar photospheric magnitude V and then applying the bolometric correction to derive the bolometric magnitude. Alternatively, stellar photometry in several bands can be used to build the SED, 
and then, by fitting it to a certain model, the stellar temperature and luminosity can be predicted. This simple approach is not possible in most cases
because of the dust in the interstellar medium, which reddens the spectral distribution and converts the determination of parameters by fitting with a model into a degenerate problem between temperature and extinction. 

We assumed
as valid the temperature values obtained from the literature with
the method explained in the previous section. We then used
a procedure that allowed us to obtain the luminosity by fitting
the SED, introducing the total extinction necessary to obtain the
already known temperature value within a range of 250 degrees, which
can be considered an acceptable value for the errors of the temperatures assigned from the literature to each object. 

The VOSA software is the Spanish Virtual Observatory tool that was designed, among other uses, to estimate effective temperature (Teff), gravity, and luminosity based on stellar photometry. The user provides the coordinates of the source, the source distance, and its uncertainties, and the system searches for observed flux (and their errors) by querying several photometric catalogues accessible through VO services to achieve as wide a wavelength coverage of the data to be analysed as possible. We were then able to choose among different stellar models to perform the fitting. We chose Kurucz models (Castelli \& Kurucz, 2003) because they are well fitted for our range of temperatures and the evolutionary stage of post-AGB stars. The VOSA software then performed the absolute flux calibration of the observational data, using the information for the available filters (zero points, transmission curves, etc.). 

Next, the software determines the synthetic photometry for the models with physical parameters in the range selected by the user (in our case, log[g] values between 0 and 5 and a metallicity between -4 and 0.5). Dust extinction is also an input to the system. It is provided without uncertainty together with the selection of an appropriate extinction law. The VOSA tool makes use of the extinction law by \citet{1999PASP..111...63F} that was improved by \citet{2005ApJ...619..931I} in the infrared. Next, the best-fitting model is provided by VOSA, together with the derivation of the corresponding stellar parameters: Teff and luminosity (and a value for log[g], the metallicity, and the overabundances of $\alpha$-elements with respect to iron). 

In the problem that interests us, we carried out an iterative procedure that consisted of providing an input test value of the extinction (starting with a value close to the interstellar extinction value from 3D maps) and determining the temperature value that was obtained in the SED fitting. We then modified the extinction value in steps of 0.05 magnitude until a temperature value closer to the literature value within the uncertainty of 250 K mentioned before was obtained. The total extinction values obtained from the SED fit can be compared with those from the extinction maps to derive the contribution of the circumstellar component.

%Interstellar extinction values
%from 3D dust maps allow for checking the consistency of the derived total extinction (interstellar plus circumstellar) coherently.
%Interstellar extinction values from 3D dust maps allow for checking the consistency of the derived total extinction (interstellar plus circumstellar) coherently.
%Concretely, we checked that the obtained total extinction values were equal to or higher than their corresponding interstellar values obtained from the dust maps.}

%\begin{figure}[h!]
%        \centering
%        \includegraphics[width=8.5cm,height=5cm]{images/HD_56126_SED.png}
%        \includegraphics[width=8.5cm,height=5cm]{images/VV399_Cyg_SED.png}
%        \caption{SED fitting for two different post-AGB candidates.}
%        \label{fig:SED}
%\end{figure}

%SED fitting was done using the Virtual Observatory SED Analyser tool (VOSA) from the Spanish Virtual Observatory (SVO) platform\footnote{http://svo2.cab.inta-csic.es/theory/vosa/index.php}. The input parameters in VOSA are celestial coordinates, distance (with uncertainty), and the total dust extinction value.
%necesitamos justificar que el ajuste de VOSA, a un único BB es suficientemente bueno, porque por ejemplo Vickers y Kamath usan varios BB para ajustar simultaneamente las envolturas o discos de polvo y la fotosfera. ¿Podría esto explicar que nuestras L sean menores que las de Outmaier y Kamath?
Post-AGB candidates in the Torun catalogue were identified as objects with an infrared excess due to the presence of a dust envelope or a disc. This means that this infrared excess should be accounted for when fitting the SEDs. In VOSA, the excess is detected by iteratively calculating (adding a new data point from the SED at a time) in the mid-infrared
(wavelengths redder than 2.5 $\mu\tab[0.01cm]m$) the $\alpha$ parameter as defined
in \citet{2006AJ....131.1574L}. The theoretical spectral models used by VOSA are based on stellar atmospheres. As a result, the tool only considers the data points of the SED to calculate the fit errors that correspond to bluer wavelengths than the wavelength in which the excess has been flagged.

The VOSA service can be used to build SEDs by querying a large variety of photometric surveys available on the platform. In general, we found fluxes for our objects from several catalogues that covered the ultraviolet, visible, and near-infrared wavelengths. The most common flux sources we used are 2MASS\footnote{Two Micron All-Sky Survey}, DENIS\footnote{Deep Near Infrared Survey}, IRAS, spaci, WISE\footnote{Wide-field Infrared Survey Explorer}, Tycho, Paunzen, UBV\footnote{Ultraviolet-Blue-Visible}, \textit{Gaia} DR3, \textit{Gaia} XPy, Pan-Starrs, and GALEX. The flux ranges are shown in the figures in Appendix B\footnote{https://zenodo.org/uploads/11569760}, and the individual fluxes for each object are listed in Table A.6 and Table A.7 (only available at the CDS). The  SEDs we provide were analysed to check for bad data points and were then fitted to Kurucz models  \citep{2003IAUS..210P.A20C}. We found these models very well suited for our sample because they cover a wide temperature range, from 3500 K to 50,000 K. 

This procedure has allowed us to derive coherent pairs of temperature/total extinction and the luminosity for each of the 118 stars in our final working sample. The uncertainty values for the luminosity were estimated by VOSA using the uncertainties in the photometry and taking into account the distance uncertainties (lower and upper limits) that we provided as input. The VOSA software provides uncertainties in luminosity  below 10\% in general, which can be considered as a lower limit. As mentioned, VOSA does not support the use of errors in the extinction values to compute the SED fitting. More information about the fitting procedure is provided in  \citet{2008A&A...492..277B} and in the VOSA documentation \footnote{http://svo2.cab.inta-csic.es/theory/vosa/}. 

This procedure, together with  accurate distances from \textit{Gaia} parallaxes, allowed us to obtain values for the luminosities and the total extinctions that agree, for instance, with those given by \citet{2022ApJ...927L..13K}  
as illustrated in Table \ref{table:kamath}. However, the fitting procedure we used to estimate the luminosity and temperature values has its own limitations because it does not take the uncertainty of the total extinction values into account (which affects the luminosity uncertainties), and it also depends on the fitting models.
%As we mentioned in Section 3.1, $\Delta A_{V}$ values are quite often below 0.15 and this translates into quite negligible differences in the estimation of the luminosity values (lower than 0.03 dex) when considering either $A_{V} + \Delta A_{V}$ or  $A_{V} - \Delta A_{V}$ for the fitting. 

%\textcolor{red}{(Minia) Kamath dice que las postAGB deben ser metal-poor, y con valores bajos de logg y tiene razon. Iker, porfa calcula las distribuciones de metalicidad y de logg que obtenemos con VOSA, espero que la mayoría de los valores ajustados sean bajos en ambas variables..y entonces los podemos comentar aqui. Hay que aislar los objetos con metalicidades por encima de 0 y con gravedades por encima de 3...y ver esos ajustes}
The VOSA tool fitting of Kurucz models also allowed us to obtain tentative values of log[g] (which covers a range between 0 and 5) and metallicity (which covers a range between -4 and 0.5) for our sample stars, although more reliable values for these parameters can be obtained from spectroscopy when available. %For the 35 objects with galactic height beyond 1250 pc (see Fig. 8, Section 3.4 and subsequent discussion), which were found to be in the Halo, we performed VOSA fittings by setting the metallicity value to [Fe/H]=-1.5, an average value for Halo stars. We checked the bibliography for references to Halo membership, and confirmed that Aoki et al. (2022) 
In Appendix B we provide the fitted SEDs for all the 118 objects in our final sample, including these parameters. The luminosity values together with lower and upper uncertainties, with the limitations explained before, are presented in Tables \ref{table:postAGB}, \ref{table:postRGB}, \ref{table:YSO} and \ref{table:supergiant}. %From now on we will refer to these objects as the final sample. 
%Figure \ref{fig:SED} illustrates some examples of SED fittings.

Following \citet{2015MNRAS.454.1468K}, we analysed the shape of the SEDs and provide a classification into three different types (stellar, shell, or disc). This can give us some additional clues about the possible incidence of binarity in our sample. According to these authors, disc-type SEDs are related to binarity. We classified 30 SEDs as disc-type. The SED morphological classification for our objects is shown in Tables \ref{table:postAGB}, \ref{table:postRGB}, \ref{table:YSO} and \ref{table:supergiant}. We opted to locate the disc-type SEDs in an HR diagram along with the remaining objects as their identification as binaries is tentative and not confirmed
in general.

\begin{figure}[h!]
        \centering
        \includegraphics[width=8.5cm,height=6cm]{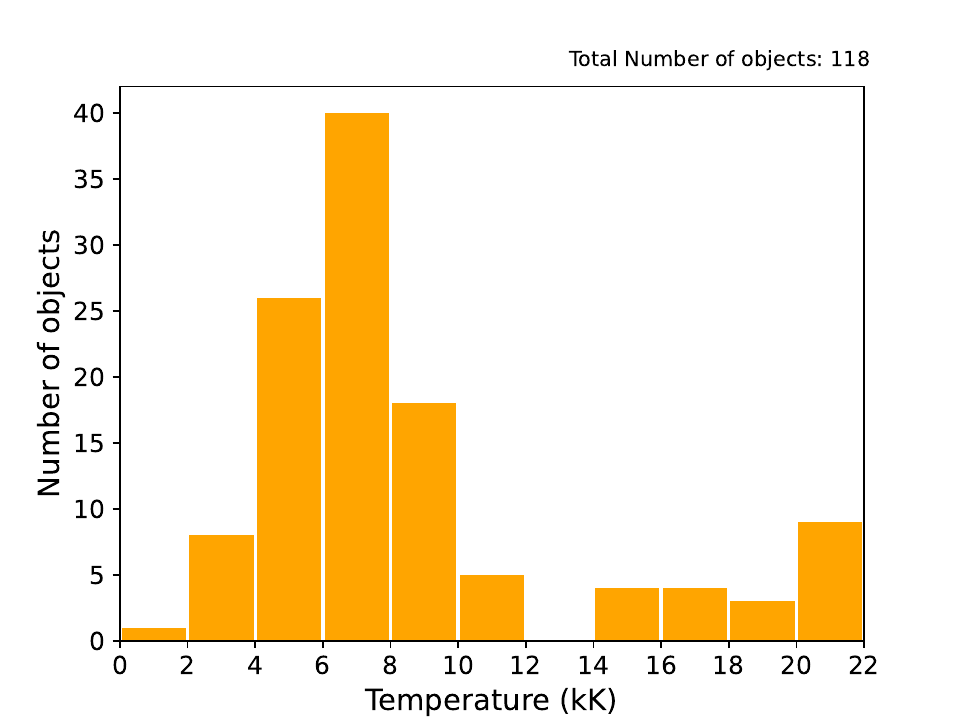}
        \caption{Temperature distribution for the 118  stars for which we provide an SED fitting.}
        \label{fig:temperature}
\end{figure}

Figure \ref{fig:temperature} depicts the temperature distribution for our final sample of objects. Most stars (84\%) have values below 10 000 K, as expected for post-AGB stars, and only three stars exhibit effective temperatures above 20 000 K. These last three are sources with infrared flux excess that have already started to ionise their envelopes on their way to the planetary nebula phase.

\begin{figure}[h!]
        \centering
        \includegraphics[width=8.5cm,height=6cm]{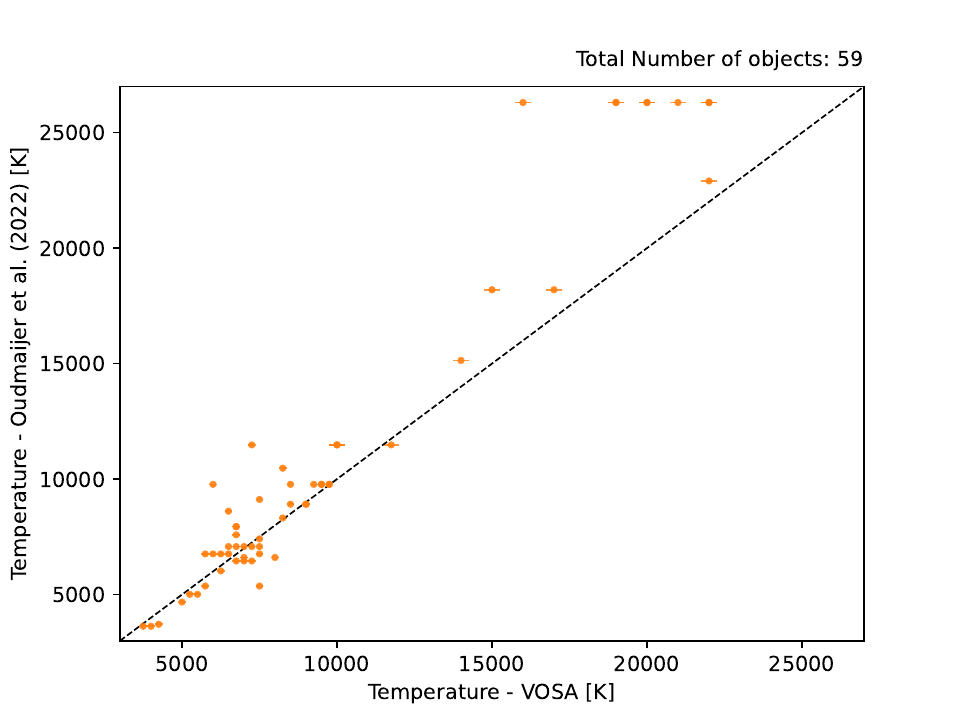}
        \caption{Temperature values from our VOSA analysis vs. those from \citet{2022MNRAS.516L..61O} for the 59 objects in common with known temperature values.}
        \label{fig:temperature_oud}
\end{figure}

When we compare our temperature determinations with those obtained by \citet{2022MNRAS.516L..61O} for the 59 objects in common in both samples, we find (Fig. \ref{fig:temperature_oud}) that the temperatures in \citet{2022MNRAS.516L..61O} tend to be slightly higher than those we obtained. The mean difference is 
$<\Delta T_{eff}> = 1.55 \pm 0.78 \hspace{0.1cm} kK.$

The temperatures in \citet{2022MNRAS.516L..61O} are based on spectral types collected from the Simbad database, which have very different origins and qualities. This might explain the discrepancies.
%{\bf not clear what a linear fit means in this plot, can you just measure the mean of the difference between both determinations? }A linear fit to the measurements show a bias that goes from $\Delta log(T_{eff})=0.05$ for lowest temperatures to $\Delta log(T_{eff})=0.2$ for the highest ones. As \citet{2022MNRAS.516L..61O} estimate temperatures from the spectral types, collected mainly from Simbad database, we assume that VOSA temperatures are more reliable. 
%In fact, these authors listed several stars with O and Wolf-Rayet spectral types, too hot to correspond to post-AGB objects and with some of them already identified as planetary nebulae in the literature.%{\bf I would say: The temperatures in \citet{2022MNRAS.516L..61O} are systematically higher than the ones obtained here and we consider the VOSA estimates to be more reliable.}

\begin{figure}[h!]
        \centering
        \includegraphics[width=8.5cm,height=6cm]{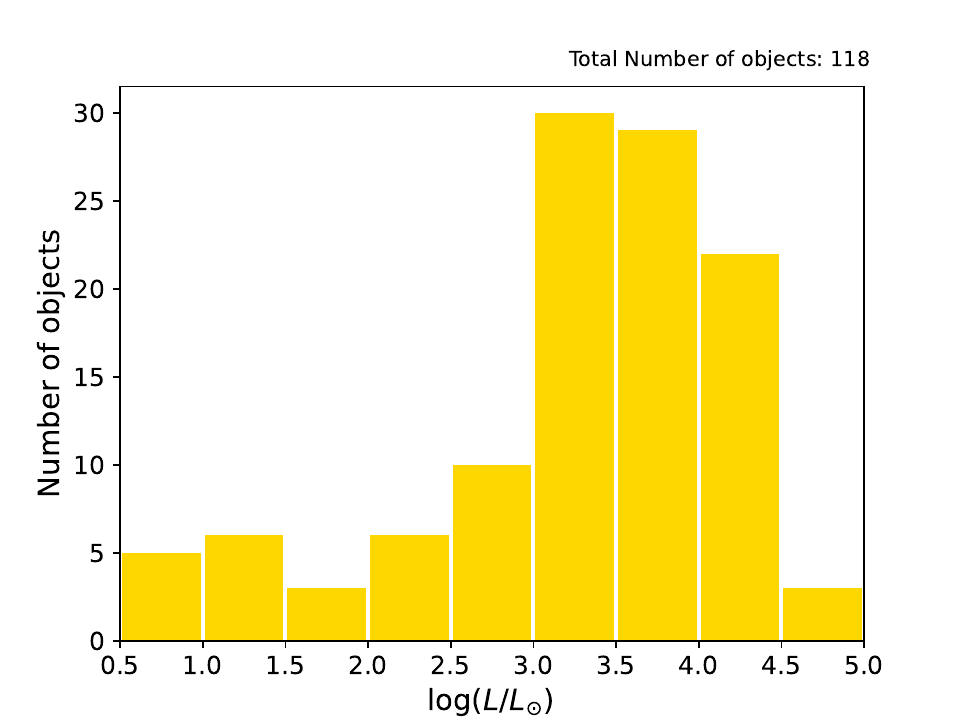}
        \caption{Luminosity distribution for the 118 stars in the final sample.}
        \label{fig:luminosity}
\end{figure}

Fig. \ref{fig:luminosity} shows that the stellar luminosities of most of the candidate objects  % , the least luminous star show less than 1 $L_{\odot}$ ($\log[\frac{L}{L_{\odot}}]=0$), while the most luminous one reach more than 100,000 $L_{\odot}$ ($\log[\frac{L}{L_{\odot}}]=5$).
lie between $2.5<\log[\frac{L}{L_{\odot}}]<4.5$. This region includes the main luminosity range expected
for post-AGB stars, but the histogram includes a wide zone of underluminous objects as well. %Concretely, both types of stars overlap in the range $3.0<\log[\frac{L}{L_{\odot}}]<3.5$. %is in which both type of stars populations overlap. 
%This is a very interesting region and it contains about 34\% of our sample.

\begin{figure}[h!]
        \centering
        \includegraphics[width=8.5cm,height=6cm]{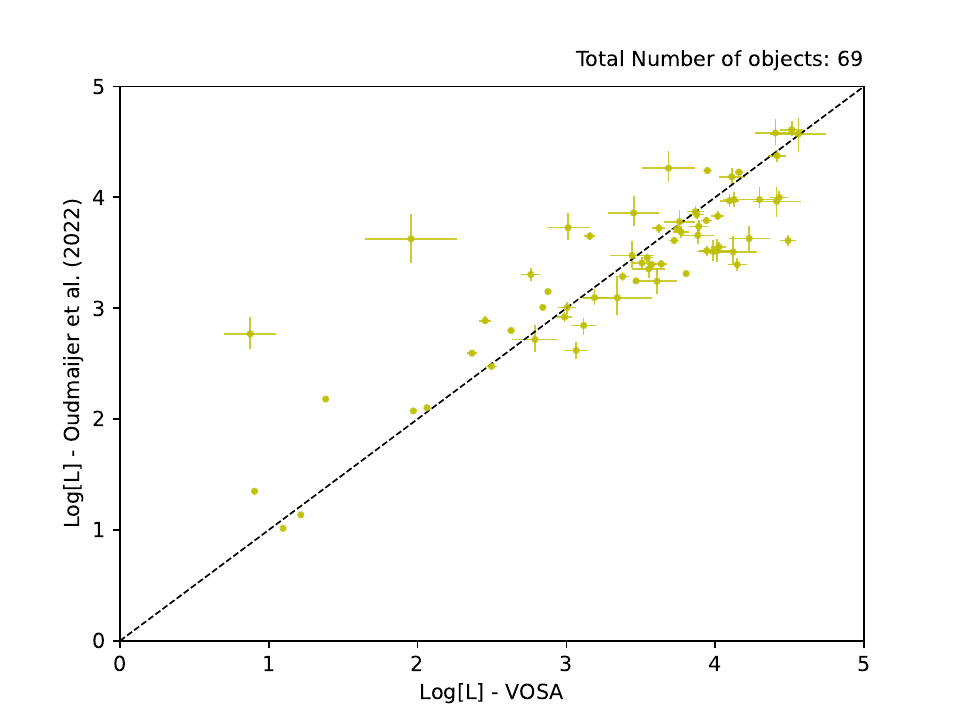}
        \caption{Luminosity values from VOSA vs. those from \citet{2022MNRAS.516L..61O} for the 69 objects in common with known luminosity values.}
        \label{fig:luminosity_oud}
\end{figure}

Figure \ref{fig:luminosity_oud} shows the comparison between our luminosity
%ies 
values and those presented in \citet{2022MNRAS.516L..61O} for the 69 stars in common. %The luminosity values from Oudmaijer tend to be overestimated comparing with those obtained from VOSA, as can be appreciated through the linear fit of the graphic. {\bf why you don´t just say that 
Not all objects with a luminosity value in \citet{2022MNRAS.516L..61O} have a temperature value.
The luminosity values from \citet{2022MNRAS.516L..61O} are very similar to those obtained in this work,
%The bias is about $\Delta \log[\frac{L}{L_{\odot}}]=1$ for the less luminous objects, while it is almost null for most luminous ones, where the agreement is quite good. Note that
 the mean difference is $<\Delta Log(L)> = 0.02 \pm 0.17.$

\citet{2022MNRAS.516L..61O} determined their luminosities through the dereddened integrated fluxes obtained from \citet{2015MNRAS.447.1673V} and by multiplying by the square of the distances from \textit{Gaia} DR3 parallaxes. The authors indicated that the errors in the fluxes are about 20\%, which could explain some differences. It is also important to note that Vickers et al. obtained their integrated fluxes assuming default values for the luminosity, which implies then that the values in \citet{2022MNRAS.516L..61O} were calculated using a rather circular argument. %As we use a more restrictive filtering constrains in parallax error to select our sample, we expect to have more accurate distances, in general, and consequently, more accurate luminosity values.

The temperature and luminosity individual values for all of our objects are available in Columns 6 and 8 of Tables \ref{table:postAGB},  \ref{table:postRGB}, \ref{table:YSO}, and \ref{table:supergiant}. 
\subsection{Galactic heights and membership to the halo}

\begin{figure}[h!]
        \centering
        \includegraphics[width=8.5cm,height=6cm]{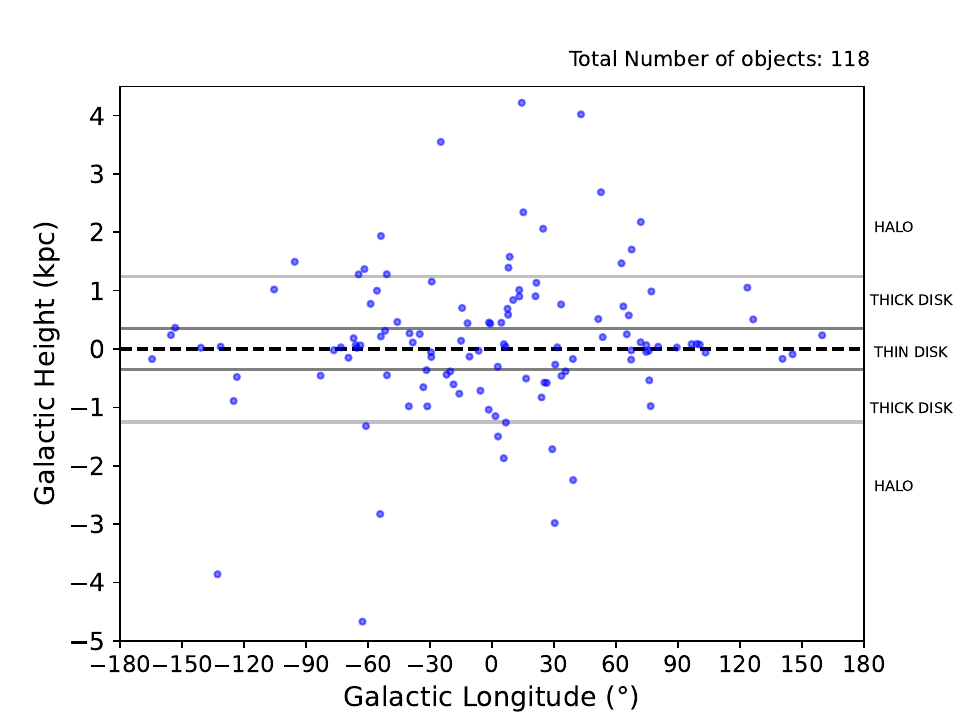}
        \caption{Galactic height vs. Galactic longitude for all 118 post-AGB final sample candidates.}
        \label{fig:gal_height}
\end{figure}

The precise \textit{Gaia} DR3 distances allowed us to calculate the Galactic distribution of our 
%general 
final sample. Figure \ref{fig:gal_height} depicts the Galactic height as a function of the Galactic longitude. It also shows the commonly adopted limits for the main structures in the Milky Way: the thin disc, thick disc, and the halo. 

%So, this also indicates a similarity between both distributions. 

This distribution allowed us to tentatively assign the 118 objects to either the 83 disc objects with $\textit{z}\leq{1.25}$ kpc) or to the halo (35). Suspected halo stars are flagged with 'H' in Tables A2, A3, and A5. Although this classification was adopted to compare their position in the HR diagram with evolutionary tracks suited for each of the populations, we are well aware that it will benefit from a spectroscopic confirmation.  

%\subsection{Identification of post-AGB stars and other objects}
\subsection{Classification: Identification of post-AGB stars and other objects}
After the luminosity values and their uncertainties were known, we applied the luminosity thresholds discussed before. We obtained that 69 objects can be classified as post-AGB star bona fide candidates, % (\textcolor{red}{36} likely and 11 possible), 
46 objects cannot be confirmed as post-AGB stars because we derived 
%for them 
a luminosity lower than 2500 $L_{\odot}$ for them, and 3 objects above the high 35,000 $L_{\odot}$ luminosity threshold are classified as supergiant stars. 

In the sample of 46 stars with luminosities lower than 2500 $L_{\odot}$, 5 are found to be YSOs in molecular clouds (see below), 3 are suspected or confirmed to be Horizontal Branch (HB) stars, 
%12 objects are YSOs according to their IRAS colors, 
and 38 remain unclassified. In this last group, 9 objects with luminosity lower than 100 $L_{\odot}$ are tentatively classified as possible YSOs, as we discuss below. %Note that all objects with at least 2500 $L_{\odot}$ ( $\log[\frac{L}{L_{\odot}}]=3.4$), considering also the error bars, have been catalogued as post-AGB stars.  While, four objects with luminosities above the threshold of $\log(\frac{L}{L_{\odot}})>4.5$ are suspected to be supergiant stars. 
The properties of the objects in the main categories are listed in Table A.2 (post-AGB stars), Table A.3 (unconfirmed post-AGB candidates), Table A.4 (YSOs candidates), and Table A.5 (supergiants and HB stars).

\begin{figure*}[h!]
        %\centering
        %\sidecaption
        \includegraphics[width=12cm]{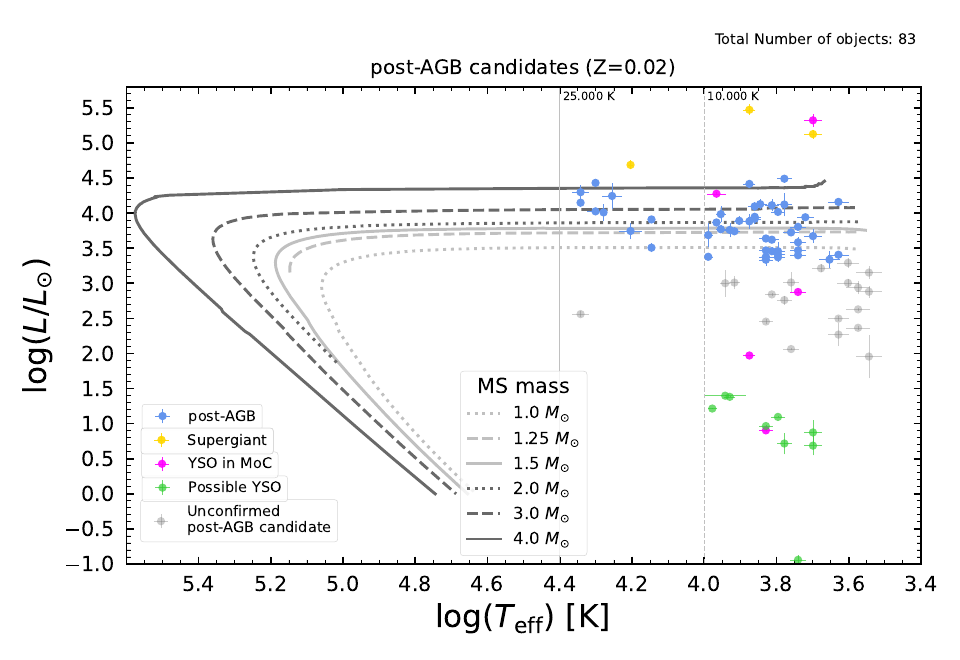}
        %\caption{Location in the HR Diagram of the 118 post-AGB candidates with luminosities and temperatures derived using VOSA. Evolutionary tracks by \citet{2016A&A...588A..25M} are shown ($Z=0.02$) and the objects are colour-coded according to the classification shown in the legend. In this diagram those objects located within the galactic disc ($\textit{z}\leq{1.25}$ kpc) are shown.}
        \\\\
        \sidecaption
        \includegraphics[width=12cm]{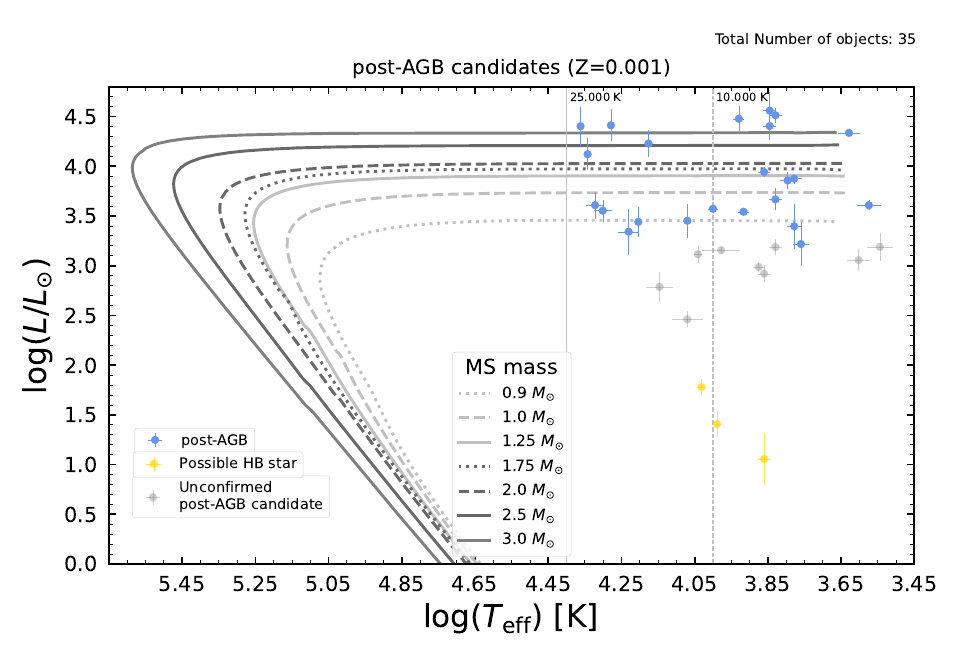}
        \caption{Location in the HR Diagram of the 118 post-AGB candidates with luminosities and temperatures derived using VOSA. Evolutionary tracks by \citet{2016A&A...588A..25M} are shown and the objects are colour-coded according to the classification shown in the legend. The upper panel shows those objects located within the Galactic disc, with $\abs{z}\leq{1.25}$ kpc, and $Z=0.02$ evolutionary tracks for comparison, while in the lower panel, those with $\abs{z}>{1.25}$ kpc (Galactic halo) together with $Z=0.001$ tracks are shown.}
        \label{fig:HRD}
\end{figure*}

To illustrate these results, we depicted all these objects in the HR diagram (Fig. \ref{fig:HRD}) together with the evolutionary tracks for post-AGB stars and PNe central stars from \citet{2016A&A...588A..25M}. To derive the masses, we used the tracks with a metallicity of $z=0.02$ for stars with $\textit{z}\leq{1.25}$ kpc that are
%HZ<1.25 Kpc, 
expected to be in the disc (upper panel), while for those with $\textit{z}>{1.25}$
kpc, we used the $z=0.001$ tracks. With the luminosity threshold for post-AGB stars discussed before, the diagram allowed us to disclose bona fide post-AGB candidates from those that are not. %All post-AGB stars (defined according to Kamath's limits) fall into the region corresponding to post-AGB stars in Miller-Bertolami tracks, within their uncertainty bars. While supergiant stars are located above the tracks, and post-RGB stars and YSOs are located below them. So this result reinforces the luminosity criteria used for identifying post-AGB stars.

%In Fig. \ref{fig:HRD}, upper panel, we distinguished between likely and possible post-AGB stars. This is to take into account those objects with $3.4<\log[\frac{L}{L_{\odot}}]<3.5$, which will fall slightly below the evolutionary tracks, but with a luminosity within the threshold used by \citet{2015MNRAS.454.1468K}. 
We would like to stress the fact that we applied a quite restrictive selection threshold to sort out bona fide post-AGB candidates. Some other objects that are located close to our luminosity threshold might also be post-AGB, but given the uncertainties, they do not fulfil our selection criteria for possible post-AGB stars.

Twelve of the objects that we classified as post-AGBs have been identified as possible or confirmed binary stars by \citet{2019A&A...631A.108K}. We checked the bibliography for other binary references and found 5 additional binary stars. We identify these stars, together with two possible YSO binary stars, also according to \citet{2019A&A...631A.108K}, one supergiant star, and one post-AGB unconfirmed candidate with a flag in Table \ref{table:candidates}. %\ref{table:postRGB}, \ref{table:YSO} and \ref{table:supergiant}.
%Tables \ref{table:postAGB} and \ref{table:YSO}. 
Moreover, our SED classification is disc-type for all but one of these binary objects. 

Figure \ref{fig:HRD} also shows two types of YSOs.  All objects with $L< 100 \hspace{0.1cm} L_{\odot}$ are suspected to be YSOs, but they might also be other types of evolved stars, such as HB stars. There are also examples of well-known YSOs among the objects with higher luminosities. The physical nature of YSOs of these luminous objects is more difficult to discern because their luminosities and temperatures overlap with those of post-AGB stars. We studied their locations in the Galaxy.  If their positions and distances matched those of star-forming regions, they were quite safely classified as young objects.

We used the molecular cloud catalogue by \citet{2020A&A...633A..51Z}, which gives coordinates and distances to a large number of these regions. We found that five of our objects are located within these clouds. These objects are labelled "YSO in MoC" (molecular clouds) in Figure \ref{fig:HRD} and Table A.4. 

The luminosities of three of these objects are above $L< 100 \hspace{0.1cm} L_{\odot}$, and the luminosities of two others are  below this limit. The classification as a YSO of the remaining eight objects with luminosities below $L< 100 \hspace{0.1cm} L_{\odot}$ is tentative. We therefore label them possible YSOs in Fig. \ref{fig:HRD}.  

Figure \ref{fig:HRD} also shows three objects that we found to be Horizontal Branch stars: SDS2012 Ter8 38 is a blue Horizontal Branch star in the globular cluster Terzan 8, [SDS2012] NGC 6402 160 in NGC 6402 and BPS BS 16479-0009 is a field Horizontal Branch star candidate according to \citet{1996ApJS..103..433B}.  Finally, we found three objects above the upper luminosity limit expected for post-AGB stars that we classify as supergiant stars. We comment further on them in section 5.

\subsection{Comparison with other classifications in the literature}

We can compare our classification with the classification recently obtained by \citet{2022PASJ...74.1368A} for the seven objects in common in both samples. Four post-AGBs are identically classified by both of us, while two of our unconfirmed post-AGB stars were catalogued as post-AGB or as cool post-AGB by them, and one of our possible YSOs was catalogued as a hot subdwarf by these authors. 

%One of these YSOs, PHL 1580, presents magnitudes and colours in Gaia DR3 that are very discrepant with the parameters reported by these authors. DR3 values are G=12.18 and  $G_{BP}-G_{RP}=0.87$, and \citet{2019yCat.4038....0S} reports a rather low extinction value of $A_{V}=0.10$. Correspondingly to those values, VOSA fluxes at the star coordinates can be fitted to a source with a temperature ($T_{eff}=5750$ K) and a luminosity ($L=1.02 \tab[0.1cm] L_{\odot}$) very similar to that of the Sun. In contrast, \citet{2022PASJ...74.1368A} reported a much %more hot 

%hotter and more luminous source with

%$T_{eff}=24,000 \tab[0.1cm] K$ and  $L=13.2 \tab[0.1cm] L_{\odot}$.
 
%\citet{1991MNRAS.248..820C} classified PHL 1580 as a high Galactic latitude
%post-AGB at a distance of 4 kpc. However, Gaia DR3 has shown that the
%actual distance is 314 pc at a Galactic height of 213 pc, therefore it is probably a
%young stellar object. It is worth noticing that in Gaia DR3 
%this source appears with the \textit{duplicate\_source} parameter as \textit{true}, possibly indicating the presence of a binary companion closer than 0.18 arcsec (Gaia minimum angular resolution). 

When comparing our results with those in \citet{2022ApJ...927L..13K}, we found 11 objects in common (those with good astrometric quality in that work, RUWE<1.4). 
%From those we 
We can confirm a nature as bona fide post-AGB candidates for 10 of these, while HD 107360 remains slightly underluminous for the post-AGB threshold. %\textcolor{red}{10} objects as post-AGB. %three of them as post-RGB (HD 56126, HD 107369, and V* V1401 Aql), 
In general, the temperatures and luminosity values given by \citet{2022ApJ...927L..13K}  agree with our derived values,
%considerably larger than ours, 
as ,illustrated in Table \ref{table:kamath}.

%\onecolumn

\begin{table*}%[h!]   

\caption{Temperature, total extinction and luminosity values for the 11 objects in common with \citet{2022ApJ...927L..13K}.}  
\label{table:kamath}      

\normalsize
%\small
%\footnotesize
%\scriptsize
%\kfontsize

\centering

\begin{tabular}  {l c c c c c c }    
\hline\hline 

Object & $A(V)^{(1)}$ & $A(V)^{(2)}$ & $T_{eff}^{(1)}$ & $T_{eff}^{(2)}$ & $Log[L]^{(1)}$ & $Log[L]^{(2)}$\\  
& (mag) & (mag) &  (K) & (K) &   &  \\ 
\hline 

HD 56126 & 2.0 & 1.33(*) & 7250$\pm125$ & 7485$\pm250$ & 3.94$_{-0.05}^{+0.04}$ & 3.74$_{-0.04}^{+0.05}$      \\
\textbf{HD 107369} & 0.3 & 0.22 & 7500$\pm125$ & 7533$\pm250$ & 2.99$_{-0.06}^{+0.06}$ & 2.96$_{-0.05}^{+0.04}$ 		\\
HD 133656 & 1.0 & 0.90 & 8250$\pm125$ & 8238$\pm250$ & 3.74$_{-0.04}^{+0.04}$ & 3.72$_{-0.03}^{+0.04}$      \\
HD 148743 & 0.6 & 0.34 & 6750$\pm125$ & 6728$\pm250$ & 4.51$_{-0.08}^{+0.07}$ & 4.41$_{-0.06}^{+0.07}$     \\
HD 161796 & 0.75 & 0.40 & 6000$\pm125$ & 6139$\pm250$ & 3.88$_{-0.05}^{+0.04}$ & 3.76$_{-0.04}^{+0.04}$      \\
HD 187885 & 1.8 & 1.74 & 8000$\pm125$ & 8239$\pm250$ & 3.89$_{-0.21}^{+0.14}$ & 3.85$_{-0.06}^{+0.06}$     \\
HD 235858 & 2.8 & 2.73 & 5250$\pm125$ & 5325$\pm250$ & 3.94$_{-0.04}^{+0.03}$ & 3.75$_{-0.03}^{+0.04}$     \\
IRAS 01259+6823 & 3.0 & 3.20 & 5500$\pm125$ & 5510$\pm250$ & 3.47$_{-0.04}^{+0.04}$ & 2.53$_{-0.19}^{+0.28}$\\
IRAS 12360-5740 & 2.7 & 3.10 & 7500$\pm125$ & 7273$\pm250$ & 3.88$_{-0.11}^{+0.08}$ & 3.80$_{-0.09}^{+0.10}$ \\
V* LN Hya & 1.0 & 0.93 & 6250$\pm125$ & 6393$\pm250$ & 4.02$_{-0.04}^{+0.04}$ & 4.03$_{-0.03}^{+0.04}$     \\
V* V1401 Aql & 1.0 & 1.24 & 6750$\pm125$ & 6985$\pm250$ & 3.47$_{-0.02}^{+0.02}$ & 3.55$_{-0.01}^{+0.02}$  \\

\hline

\end{tabular}

\tablefoot{\myfontsize{(*): Note the A(V) difference between both works, higher than 0.5 mag. \cite{2012MNRAS.419.1254R} did not find a C/O ratio greater than 1 nor s-process enrichment in HD 107369, as it is expected for an AGB star. In the present study, this star is classified as an unconfirmed post-AGB candidate.} }

\tablebib{\tiny{(1): calculated by VOSA, (2): \citet{2022ApJ...927L..13K}}.}

\end{table*}

%\twocolumn

%The differences between \citet{2022ApJ...927L..13K} and our luminosities are basically due to the much higher extinction values used by \citet{2022ApJ...927L..13K}. Such extinctions were derived directly from their SED fitting, while our values are taken from 3D dust maps in the literature, under the assumption that the interstellar reddening in the direction of the star represents the total reddening of the sources, in a different methodological approach. %\textcolor{red}{Note that for this analysis we have approximated the total reddening of the objects as the interstellar extinction. However, for a more accurate estimation, the circumstellar extinction of the objects should be considered. Consequently, our luminosity estimations could be slightly underestimated}.
In Section 3.3 we compared our results with those obtained by \citet{2022MNRAS.516L..61O}, as shown in Figures 5 and 7. For the luminosity range expected for post-AGB stars, the temperatures and luminosities agree with a dispersion that can be explained by methodological differences, as already discussed in that section. Forty-four of the 59 objects in common with \citet{2022MNRAS.516L..61O} sample were classified as post-AGB 44, and we catalogued 8 of them as unconfirmed candidates, 3 as possible YSOs, 2 as YSOs in molecular clouds, and another 2 as supergiants.

Finally, we summarise our results. Starting from the lists of post-AGB objects known or proposed as such in the literature, we selected stars based on the quality of the astrometry in \textit{Gaia} DR3 of these sources. We used updated dust-extinction maps, 3D when available, to derive more accurate luminosities. As a consequence, 
%we were able to locate a higher number of sources in the HR diagram (146 vs 134) and 
we were able to classify some of them as bona fide post-AGB stars (69),
supergiants (3), HB stars (3), YSOs in molecular clouds (5), and possible YSOs (9), while 29 objects remain unconfirmed post-AGB candidates. 
In the following section, we describe the evolutionary properties of our sample of 69 post-AGB stars. %and give information about some interesting individual objects. %Moreover, we have estimated the mass and the evolutionary age for the sample of our confirmed 45 post-AGBs, which are also individually available in Table \ref{table:postAGB}. 

%EVA: {\bf aqui tienes que hacer una comparación más cuantitativa para demostrar que este paper añade valor a la publicación que ya hay en la literatura. Mira sus conclusiones y determina que diferencias encuentras haciendo el analisis con VOSA, los tamaños de las muestras son similares, o dices algo aqui mas cuantitativo o el referee va a protestar.}

\section{Sample of post-asymptotic giant branch stars}

We focus now on the sample of 69 objects whose luminosities allowed us to confirm their evolutionary state as bona fide post-AGB candidates. 
By interpolating between the novel evolutionary models by \citet{2016A&A...588A..25M} for post-AGB stars, we estimated their progenitor mass (in the MS) and their evolutionary age in the post-AGB phase. \citet{2016A&A...588A..25M} provided tracks for 
%metallicities 
metallicity values 0.01, 0.02, 0.001, and 0.0001. We used the tracks for $Z=0.02$ as 
%more 
representative of the disc population, and for objects belonging to the halo, we used $Z=0.001$ tracks. 

%\textcolor{red}{Then, by interpolating the location of these objects in the HR diagram between Miller-Bertolami tracks, we obtained their masses and ages. Remember that these objects have metallicities that go from -4 to 0.5, and there is not any Miller-Bertolami model which covers all such metallicities. So this procedure can only be used to have an approximation of the masses and ages.
%Furthermore, these estimations have uncertainty values that depend strongly on the luminosity and temperature values, which could be quite large in several cases.}
The objects classified as unconfirmed post-AGB candidates are located below the 1 $M_{\odot}$ track  (0.9 $M_{\odot}$ for halo stars). It is assumed that the initial masses of these objects, if they are single-evolved stars, can only be slightly below 1 $M_{\odot}$. Conversely, they could have their origin in binary evolution.

%\begin{figure}[h]
%        \centering
%        \includegraphics[width=8.5cm,height=6cm]{images/postAGB.pdf}
%        \caption{Location in the HR Diagram of the \textcolor{red}{48} post-AGB stars, together with evolutionary tracks by \citet{2016A&A...588A..25M}.}
%        \label{fig:HRD_2}
%\end{figure}

%EVA:{\bf tal y como has pintado el histograma parece que la mayor parte son de masas menores que 1 lo cual sería un problema porque no han tenido tiempo de evolucionar a post-AGBs teniendo en cuenta la edad del universo. yo haria otros bins en el histograma. 
%--explica como calculas la masa de las que estan por debajo de las trazas de 1Msun
%--yo daria la masa promedio de las post-agb tambien y luego la trasformas en la masa inicial....
%--añade un parrafo comparando con la distribución de masas de las estrellas centrales de nebulosas planetarias en la galaxia  
%https://ui.adsabs.harvard.edu/abs/2016A%26A...593A..29M/abstract
%aqui damos una masa de 0.59msun }

\begin{figure}[h!]
        \centering
        \includegraphics[width=8.5cm,height=6cm]{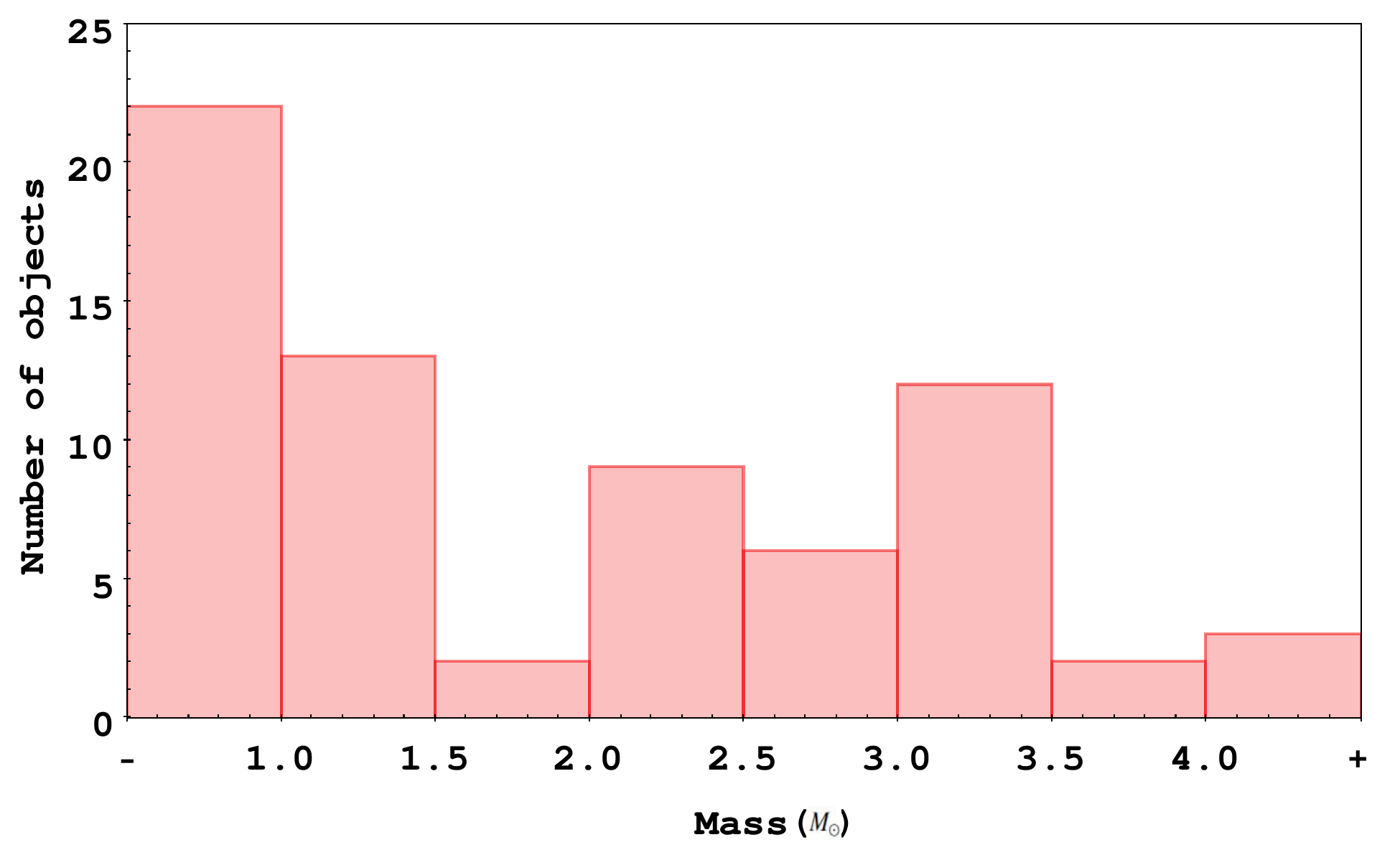}
        \caption{Progenitor mass distribution for the 69 post-AGB objects.}
        \label{fig:mass}
\end{figure}

%As a result we obtained 
The progenitor mass distribution we obtained for the post-AGB stars in our sample is displayed in Fig. \ref{fig:mass}. The masses of about half of our stars (35) are below 1.5 $M_{\odot}$ in the MS, and the masses of only 5 of them are higher than 3.5 $M_{\odot}$. Although our study is limited in the number of objects and possibly comes from a biased selection, the resulting masses match the expected distribution. The lifetimes of parent stars with masses above 3.5 $M_{\odot}$ are too short in the post-AGB phase (the crossing times for post-AGB and PNe phases in Miller Bertolami tracks are shorter than 400 yr) for them to populate this region. %, while those with masses around $\approx$ 1 $M_{\odot}$ might evolve so slowly that their envelope can be expected to thin out and become more difficult to detect. %So, we can conclude that, in general, our post-AGB objects are low-mass stars. We have obtained the following mass mean value of the sample:
The mean value of the progenitor masses for the post-AGB sample is

$$<M>_{MS} = 1.94 \pm 0.53 \hspace{0.1cm} M_{\odot}.$$

This mean mass value agrees with the value of $1.8 \pm 0.5 \hspace{0.1cm} M_{\odot}$ obtained in \citet{2021A&A...656A..51G} for stars in the next evolutionary phase, as central stars of planetary nebulae. %which has a value of $1.8 \pm 0.5 \hspace{0.1cm} M_{\odot}$. 

%The mean value of the mass of the post-AGB stars resulted: 
The mean mass value for the post-AGB stars final sample is
$$<M>_{postAGB} = 0.598 \pm 0.163 \hspace{0.1cm} M_{\odot}.$$

\begin{figure}[h!]
        \centering
        \includegraphics[width=8.5cm,height=6cm]{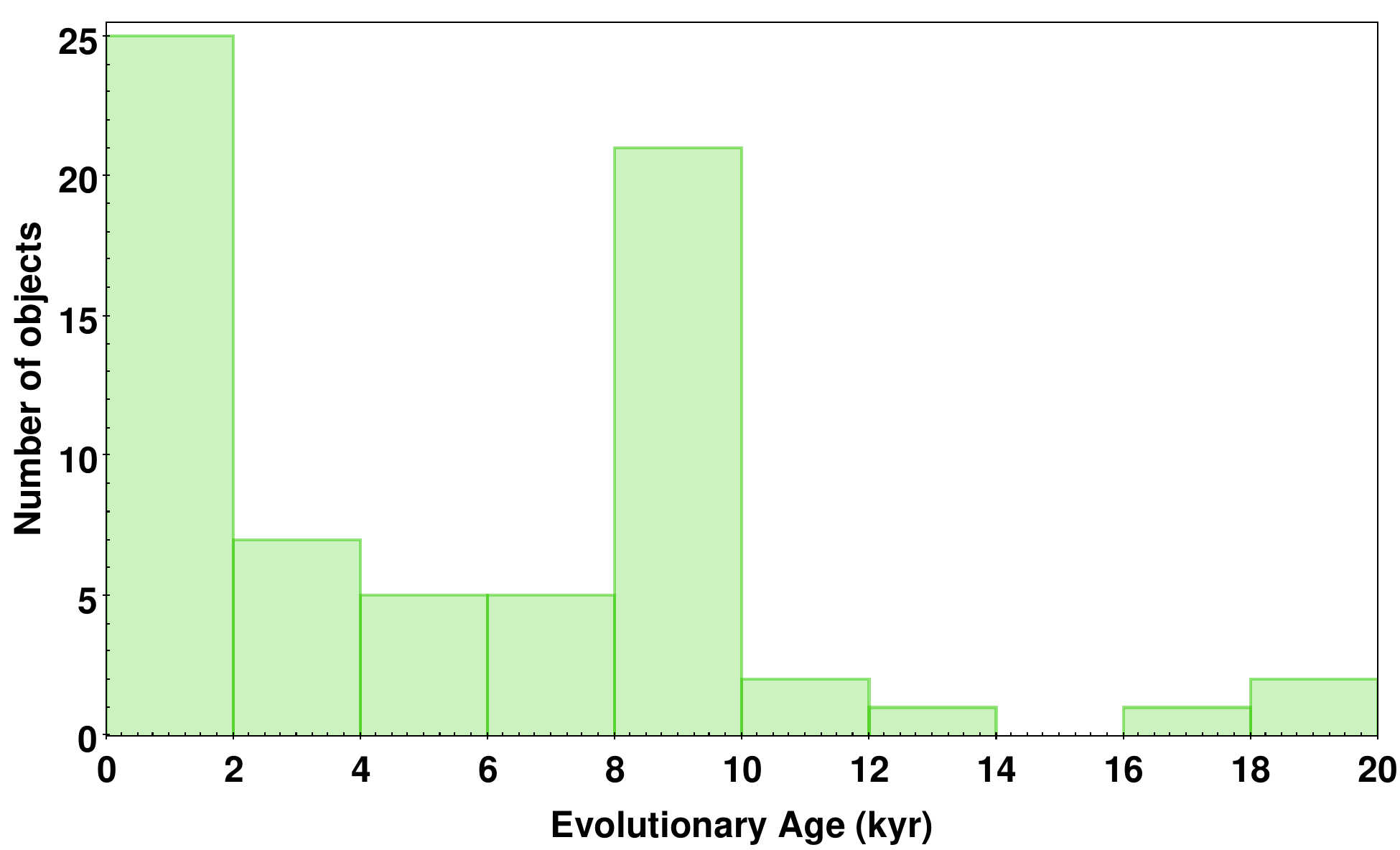}
        \caption{Distribution of the evolutionary age for the 69 post-AGB objects.}
        \label{fig:age_evo}
\end{figure}

We obtained the evolutionary age distribution shown in Fig. \ref{fig:age_evo}. The ages of about 36\% of the stars are younger than 2000 yr. Moreover, almost all the objects in the sample show evolutionary ages below 10,000 yr, and only 6 stars are older.
%in our sample. %According to this, we can corroborate that stars in the post-AGB phase evolve to the PNe phase in a very short time, of the order of a few thousand years. 
We obtained the following mean evolutionary age of the sample:

$$<T>_{evo} = 5.66 \pm 3.60 \hspace{0.2cm} kyr.$$

In Miller Bertolami models, the beginning of the post-AGB phase is taken when the mass of the external layer of the star drops below 
%the 
1\% of the star mass as a result of stellar winds.

%It is interesting to mention in this context the high relative number of stars in the post-RGB phase that we obtain in our analysis. This highlights the importance of understanding the evolution of binary systems in advanced evolutionary phases in order to correctly interpret HR  diagrams.

Individual mass and evolutionary age values can be found in Columns 11 and 12 of Table \ref{table:postAGB}, respectively.

%EVA: {\bf yo añadiria una discusion de la fraccion de objetos en esta etapa evolutiva comparado con las enanas blancas y las planetarias. El alto numero de post-RGB stars implica evolucion en sistemas binarios para una fracción elevada de estos objetos}

%As we know, post-AGB stars will evolve in CSPNe and finally in White Dwarfs (WDs). So, it is interesting to see which is the proportion of post-AGB stars over CSPNe/WDs. In a similar study that we carried out over the later ones, \citet{2021A&A...656A..51G}, we were able to study the evolutionary state of a total of 74 CSPNe/WDs, and now, we have obtained 45 post-AGB stars. So we can say that for a total of 119 objects in these evolutionary states, around the 38\% of them are in post-AGB phase, will the remaining 62\% are in CSPN/WD phase. In spite of WDs are fainter and consequently more difficult to detect, we have obtained more proportion of stars in CSPN/WD phase, this fact shows the brief lifetime of the post-AGB phase. 

%MINIA. Yo esto de los tipos espectrales no lo veo interesante. Vienen de las temperaturas..y ya se ve en la figura 9 el rango de pemperaturas, que va de 25K a 3K
%Another interesting evolutionary parameter is the spectral type. We took this values from Simbad database as well as from Torun catalogue, and we listed them in the last column of Table \ref{table:postAGB}. As can be appreciated there is a large variety of spectral types, covering almost all the spectral range. However, we can see that the majority of our post-AGB objects are F stars (33\%), A stars (22\%) or B stars (20\%).   

\section{Special objects}

Our study allows us to confirm the nature of three objects as post-AGB stars. They lacked a previous firm classification in the catalogues and were neither listed in the Torun catalogue as likely, nor in Simbad or in the Suárez et al. catalogues as confirmed post-AGB.

%As mentioned before, we catalogued as \textit{doubtful} post-AGB stars those objects which have not been confirmed before  Concretely, we have 3 \textit{doubtful} objects that we are going to discuss in detail.

% YSO in MC
%One of them is IRAS 18084-1737, the only object that falls clearly in a high temperature region (with $T_{eff}=$ 50,000 K), which corresponds to stars already evolved to the planetary nebula phase. Its 6 cm emission (\citealt{1991A&AS...91..481R}) 
%also indicates that this star must be in the planetary nebula phase. And the study of \citet{1997AN....318...35K} catalogued this object as a possible pre-planetary nebula. It was included in our sample because it appears catalogued as \textit{candidate post-AGB} in Simbad. In conclusion, it is a protoplanetary nebula and not really a post-AGB star.

CD-30 15464. This object is catalogued as possible post-AGB in the Torun catalogue but as simple star in Simbad, and it has no references in \citet{2006A&A...458..173S} catalogue. It is reported as a star of spectral type B1, while its effective temperature is quite high (22,000 K), still within the limits of a post-AGB star. It is located very far from the Sun, at a distance of 9.96 kpc, in the direction of the Galactic centre ($l = 1.67^{\circ}$) and slightly below the Galactic disc ($b = -6.63^{\circ}$). Despite this location, the interstellar extinction is low in the direction of this star ($A_{V}=0.75$), and its image\footnote{http://cdsportal.u-strasbg.fr/?target=CD-30\%2015464} in PanSTARRS colours shows evidence of a circumstellar envelope.  

%So, this location makes very difficult to distinguish this object well enough to confirm its stellar type.  

HD 53300 is catalogued as an A2 type star and is located in the Galactic disc ($b = 0.44^{\circ}$). Instead, \citet{2012MNRAS.419.1254R} derived an effective temperature of 7250 K for it, which corresponds to an F1 type star. % on the MS. 
This star is mentioned as a candidate post-AGB in the Simbad database and lacks references in the Torun or \citet{2006A&A...458..173S} catalogues, although \citet{2012MNRAS.419.1254R} classified it as post-AGB star, while it was classified in \citet{2000A&A...362..978B} as a Vega-like star. %However, the only reference we have of this object is from Simbad database, catalogued as \textit{candidate post-AGB}. 
%Gaia DR3 distances have allowed us to calculate its luminosity and confirm the classification of this object as a post-AGB star. %that this object it could be a post-AGB star, as it falls in the correspondent region of the HR diagram, according to its temperature and luminosity. Regarding other references in the literature of this object, in the study of \citet{2012MNRAS.419.1254R} it is classified as post-AGB star, while in \citet{2000A&A...362..978B} it was identified as Vega-like star. 

%EVA: {\bf  yo moveria 4.1 y 4.2 a la seccion 3 y dejaria sola la discusion de las post-agbs--añade un parrafo sobre cuantas están missclasified ahora que tienes mejores distancias}

HD 214539.This star is catalogued as possible post-AGB in Torun and as simple star in Simbad, with a spectral type B8/9. \citet{1984ApJ...278..208K}  obtained a temperature of 10,000 K and a gravity of log[g]=2 for this object. % and a very low metallicity of Fe/H=-4 for this object. 
It is located at a distance of 1.39 kpc. By visual inspection\footnote{http://cdsportal.u-strasbg.fr/?target=HD\%20214539}, a circumstellar envelope typical of post-AGB stars can also be observed.

% SUPERGIANTS

It is also interesting to analyse the three highly luminous objects ($\log[\frac{L}{L_{\odot}}]>4.5$) that we have catalogued as supergiant stars. 

BD-02 4931. This object is catalogued as a post-AGB candidate star in the Simbad database, while in \citet{2000AAS...197.6005P}, it is classified as a B1 type giant.
%and by \citet{2015MNRAS.454.1468K}, with a luminosity of $\log[\frac{L}{L_{\odot}}]=4.41$ and temperature of 6728 K. On the other hand, in \citet{2015MNRAS.446.3447L}, it is catalogued as a supergiant or bright giant star, with a temperature of 7400 K. 
Assuming a temperature of 16,000 K from its spectral type (the same value as was obtained by fitting its SED by \citealt{2011A&A...533A..99G}), we obtained a luminosity of $\log[\frac{L}{L_{\odot}}]=4.69$, which is higher than the predictions for post-AGB stars. %and surface gravity of $\log[g]=0.5$. %So it should be probably a supergiant star.

%HD 167402 -- One of them is HD 167402. It is catalogued as a post-AGB star in the Simbad database and as a hot massive post-AGB star with a spectral type of B0I and a temperature of 29,800 K in \citet{2020MNRAS.494.2117H}. Through the VOSA tool, we have estimated a temperature of 19,000 K and a luminosity of $\log[\frac{L}{L_{\odot}}]=4.68$, considerably above the threshold set for post-AGB stars. %Then, we catalogued this object as a supergiant star.% MINIA. Hayque ver esto con cuidado porque Herrero hace análisis de espectros de alta resolución y quizá sus valores sean más precisos.

HD 179821. This star is catalogued as likely post-AGB in the Torun catalogue and as a post-AGB star in the Simbad database and in the Suárez et al. catalogue. However, we obtained a high luminosity of $\log[\frac{L}{L_{\odot}}]=4.75$ (very similar to that obtained by \citet{1983ApJ...272...99W}, $\log[\frac{L}{L_{\odot}}]=4.7$) and low surface gravity of $\log[g]=0.5$ for this object, which led us to classify it as a supergiant star. Moreover, \citet{2016MNRAS.461.4071S} catalogued it as a likely massive post-red supergiant star.

V* V1027 Cyg. This object is also classified as likely post-AGB in the Torun catalogue and as post-AGB in Simbad database. In this case, however, we obtained stellar parameters typical of supergiant stars, such as a high luminosity of $\log[\frac{L}{L_{\odot}}]=4.75$ and a very low surface gravity of $\log[g]=0$. Furthermore, in the \citet{1994AAS...185.4515W} catalogue, they classified it as a G7 supergiant star, but it was catalogued as a G8-K3 type supergiant in a more recent study by \citet{2016AstL...42..756A}.

\section{Conclusions}

 Based on a sample of 118 post-AGB star candidates selected from the literature and after filtering out stars for which \textit{Gaia} DR3 astrometry was not accurate enough (and for which the distances were therefore unreliable), we have estimated their luminosity values, which 
 allowed us to classify them as bona fide post-AGB candidates (69) or as objects in a different evolutionary phase, such as YSOs (5), possible YSOs (9), supergiant stars (3), and HB stars (3). Twenty-nine stars remain unconfirmed post-AGB candidates.
 
Using the Spanish VOSA service, we fitted the SEDs of each star and simultaneously obtained its effective temperature and luminosity. This allowed us to plot the post-AGB candidates in an HR diagram, and by using the \citet{2016A&A...588A..25M} evolutionary tracks in the post-AGB phase, we derived their masses and ages. We found that our 69-object sample includes mainly stars with progenitor masses between 1 and 2.5 $M_{\odot}$, which agree with the type of post-AGB stars that statistically could be found in a small sample like ours. The mass mean value of the sample also agrees with that expected for
stars in the planetary nebula phase (the next evolutionary phase) according to \citet{2021A&A...656A..51G}.
%\textcolor{red}{Note that progenitor masses have been estimated by an specific evolutionary tracks which corresponds to an specific metallicity (Z=0.02), and that the variety of metallicities among the post-AGB stars is quite wide. So these masses are only an approximation to their real values.}

Our study allowed us to confirm the nature of several objects as post-AGB stars that were previously not confirmed as such. %objects and discard some other candidates from the compilations available in the literature. 
It is important to note that although the \textit{Gaia} DR3 catalogue contains statistically valuable information for an enormous number of stars, many of the available parallaxes contain errors that are too large to allow a correct estimation of the absolute magnitude or luminosity of the stars. We chose to perform a rather restrictive filtering of the astrometric quality of \textit{Gaia} measurements (parallax and distance errors, RUWE value) as well as of the distance inferred from the parallax measurement by a Bayesian model of Galactic stellar distributions. 

This filtering was based on the idea of working with a small set of objects, candidates for the post-AGB phase, all of which have precise distances in \textit{Gaia}, which therefore mostly are individual objects for which a luminosity can be calculated with great confidence. For this reason, our method was based on selecting only a subset of stars that are firm candidates to be in the post-AGB phase and have \textit{Gaia} DR3 parallaxes (and inferred distances) with errors below 30\%.

We also only worked with candidates with interstellar extinction values from Stassun 3D dust maps, which allowed us to initially constrain interstellar extinction. The total extinction, including both interstellar and circumstellar extinction, was derived simultaneously with the luminosity value from the SED-fitting procedure. The effective temperatures were taken from the literature, using spectroscopically determined temperatures when they were available.
%Assuming those extinction values, effective temperatures can be obtained with no degeneracy from the SED fitting of the stars. We checked the hypothesis of a low or negligible circumstellar contribution to the total extinction by considering 2MASS infrared colours of our dereddened sample and quoted the small number of cases for which circumstellar extinction can be important. This, together with the fact that our stars have all good astrometric quality, gives us confidence in using this simplification.

We discussed our results in comparison with other similar studies of post-AGBs that were carried out recently. About 25\% of the stars in our sample are too underluminous to be confirmed as post-AGB stars under our working hypotheses, while 12\%  are found to be possible YSOs, are either located in Molecular Clouds (5) or are candidates (9). Other objects, such as HB stars (3) or supergiant stars (3), were also included in the original compilation.

Although our initial filtering would rule out most of the binary objects, we found that the SEDs of 18 objects out of our sample of 69 post-AGB can be classified as disc-type. They might therefore be binary objects. Searching the literature for binarity, we found that 17 of these objects were identified as such either by Kluska et al. 2022 (12 objects) or by other authors (see table A.1 for references). This means that our sample contains 18 possible or confirmed binaries, which is about 26\%. 

Our results provide an interesting framework for further insight into the post-AGB phase. In particular, our well-characterised sample of 69 objects opens the way to complementing the study on the unconfirmed 29 candidates. A follow-up analysis of their properties, including spectroscopy when possible, would be desirable.

% we present a sample with better precision in parallaxes and distances, and this has allowed us to obtain more accurate luminosities, that consequently, allows us to classify with more reliability the objects of the sample among different stellar phases (post-AGB, YSO or supergiant). 

%It is interesting to highlight the relatively high proportion of objects we have found that appear to be in a post-RGB phase, compared to the number of post-AGB objects found. The relationship is 68/45. Despite having a rather poor statistics, this highlights the necessity of taking into account the outcomes of binary evolution to correctly interpret HR diagrams in advanced evolutionary phases.

% Moreover, we have started our analysis considering more post-AGB candidates than the other recent studies. 

%Interestingly, we have found a set of objects in the LMC direction that according to Gaia parallaxes are located out of the Galaxy but closer than the LMC, possibly in the streaming between the LMC and the Galaxy. These objects deserve further study.

\begin{acknowledgements}
This work has made use of data from the European Space Agency (ESA) \textit{Gaia} mission,  
processed by the \textit{Gaia} Data Processing and Analysis Consortium (DPAC). Funding for the DPAC has 
been provided by national institutions, in particular, the institutions participating in the 
\textit{Gaia} Multilateral Agreement. This research has made use of the Simbad database and the
Aladin sky atlas, operated at CDS, Strasbourg, France. 
The authors have also made use of the
VOSA software, developed under the Spanish Virtual Observatory project supported by the Spanish 
MINECO through grant PID2020-112949GB-I00, and partially funded by the European Union's Seventh Framework Programme (FP7-SPACE-2013-1) for research, technological development and demonstration under grant agreement no. 60674. 
%This work has been funded by the Spanish MCIN/AEI/10.13039/501100011033 and European Union Next Generation EU/PRTR through grant PID2021-122842OB-C22 and the Horizon Europe [HORIZON-CL4-2023-SPACE-01-71], SPACIOUS project funded under Grant Agreement no. 101135205.
%Funding from Spanish Ministry project PID2021-122842OB-C22, Xunta de Galicia ED431B 2021/36 and PDC2021-121059-C22 is acknowledged by the authors. 
%We also acknowledge support from CIGUS-CITIC, funded by Xunta de Galicia and the European Union (FEDER Galicia 2014-2020 Program) through grant ED431G 2019/01. 
This research was funded by the Spanish Ministry of Science MCIN / AEI / 10.13039 / 501100011033 and the European Union Next Generation programme EU/PRTR through the coordinated grant PID2021-122842OB-C22 and the Horizon
Europe [HORIZON-CL4-2023-SPACE-01-71] SPACIOUS project, Grant Agreement no. 101135205. Additionally, it is co-financed by the EU through the FEDER Galicia 2021-27 operational programme, Ref. ED431G 2023/01.
E.V. acknowledges support from the 'DISCOBOLO' funded by the Spanish Ministerio de Ciencia, Innovación y Universidades under grant PID2021-127289NB-I00. M.M. and E.V. acknowledge support from the cooperation agreement between the IAC and the Fundaci\'on Jes\'us Serra for visiting grants. AM acknowledges support from the ACIISI, Gobierno de Canarias and the
European Regional Development Fund (ERDF) under grant with reference PROID2020010051 as well as from the State Research Agency (AEI) of the Spanish Ministry of Science and Innovation (MICINN) under grant PID2020-115758GB-I00.
\end{acknowledgements}

\bibliographystyle{aa} 

\bibliography{bibliopn}

\begin{appendix}

\onecolumn

\section{Data Tables}

% TABLE 0

%\begin{center}

\scriptsize

\begin{longtable}{l c c c c c c c c c} 
\caption{General data of the 157 post-AGB candidates}.
\label{table:candidates}\\

\hline\hline
Num & Simbad Name & \textit{Gaia} EDR3 ID & RA & Dec & G mag & $A_{V}^{IS}$ & Spectral Type & Reference & Flag \\  
&  &  & (º) & (º) &  &  &  &    \\ 
\hline
\endfirsthead
\multicolumn{9}{l}
{\tablename\ \thetable\ -- \textit{Continued from previous page}} \\
\hline\hline
Num & Simbad Name & \textit{\textit{Gaia}} EDR3 ID & RA & Dec & G mag & $A_{V}^{IS}$ & Spectral Type & Reference & Flag\\  
 &  & & (º) & (º) & & &  &  &    \\ 
\hline
\endhead
\hline \multicolumn{9}{r}{\textit{Continued on next page}} \\
\endfoot
\hline
\endlastfoot     

1 & *  42 Cyg & 2056972679739120000 & 307.335 & 36.4547 & 5.74 & 1.52 & A2Iab-Ib & 1 & -                                         \\                                              
2 & *  89 Her & 4582795323914832000 & 268.8549 & 26.05 & 5.35 & 0.17 & F2Ibp & 1/2 & B(1,2,3)                \\    
3 & [SDS2012] NGC 6284   116 & 4112726164275562496 & 256.1875 & -24.55 & 16.93 & - & - & 1 & -               \\
4 & [SDS2012] NGC 6402   160 & 4368932547017571584 & 264.1583 & -3.3867 & 16.85 & 1.64 & - & 1 & -           \\
5 & [SDS2012] Ter 8   38 & 6741750932641613952 & 295.42 & -34.0664 & 15.09 & 0.41 & - & 1 & -                \\
6 & 2MASS J00235767-7205296 & 4689637789379454848 & 5.9903 & -72.0916 & 11.09 & 0.09 & - & 1 & -             \\
7 & 2MASS J01302276-7303339 & 4686479648370453760 & 22.5949 & -73.0594 & 10.63 & 0.12 & B & 1/2 & -       \\
8 & 2MASS J05241036-2429206 & 2957941232276476800 & 81.0432 & -24.4891 & 12.09 & 0.09 & - & 1 & -            \\
9 & 2MASS J06544616-1048325 & 3049274119844532224 & 103.6924 & -10.8091 & 12.05 & 2.03 & OB+ & 1 & R(1)         \\
10 & 2MASS J13272898-4722472 & 6083708479176016128 & 201.8708 & -47.3798 & 13.09 & 0.37 & - & 1/2 & R(1)     \\
11 & 2MASS J14034398-6937097 & 5846979777414443264 & 210.9333 & -69.6194 & 16.39 & 0.79 & - & 1 & -          \\
12 & 2MASS J16570924-0404243 & 4365635249084745600 & 254.2886 & -4.0734 & 13.49 & 0.79 & B & 1 & R(1)           \\
13 & 2MASS J17390218-4500388 & 5955201232284272384 & 264.7591 & -45.0108 & 13.01 & 1.15 & - & 1/2 & R(1)     \\
14 & 2MASS J17442550-1937537 & 4119884023727270272 & 266.1063 & -19.6316 & 12.43 & 1.67 & OB+ & 1 & -        \\
15 & 2MASS J18224265-3014383 & 4046476465531783424 & 275.6777 & -30.244 & 11.8 & 0.39 & B7Ib & 1/2 & -    \\
16 & 2MASS J18530579-0842378 & 4203848980711226112 & 283.2741 & -8.7105 & 13.14 & 1.14 & - & 1/2 & -      \\
17 & BD+32 2754 & 1324742534573959424 & 249.0487 & 32.4893 & 9.47 & 0.07 & F8 & 2 & -                        \\
18 & BD+33  2642 & 1369896865785991424 & 237.9995 & 32.9484 & 10.79 & 0.08 & B2 IVp & 1/2 & B(4)             \\    
19 & BD+48  1220 & 255225480926107392 & 76.9595 & 48.4026 & 9.53 & 0.89 & A4Ia & 1/2 & -                  \\
20 & BD-02  4931 & 4213102543594938880 & 289.5947 & -2.703 & 10.42 & 1.68 & B1III & 1 & B(5)                     \\ 
21 & BD-13  5550 & 6879196723703009920 & 300.4576 & -12.6883 & 11.34 & 0.44 & B1Iae & 1/2 & -             \\
22 & BPS BS 16479-0009 & 3939536182204010880 & 198.4999 & 18.5253 & 13.61 & 0.06 & - & 1 & -                 \\
23 & CD-24 13065 & 4113478337639987200 & 256.0433 & -24.4661 & 11.08 & 0.47 & B8 & 1 & -                     \\
24 & CD-30 15464 & 4049379725984965120 & 274.002 & -30.7565 & 11.9 & 0.75 & B1Ib & 2 & -                     \\
25 & CD-42  8141 & 6135778223095320960 & 197.1931 & -43.4642 & 10.45 & 0.34 & B2I & 1 & -                    \\
26 & CD-46 11775 & 5948818331093816448 & 265.6416 & -46.9802 & 11.17 & 0.61 & OB+ & 1 & -                    \\
27 & CD-48 11445 & 5938738764416910848 & 256.9027 & -48.319 & 10.5 & 2.24 & G2p(R) & 1 & B(1,5)                 \\ 
28 & CD-49  8217 & 6093466301247487488 & 207.3233 & -50.3793 & 10.82 & 0.61 & B2I & 1 & R(1)                    \\
29 & CD-49 11554 & 5946845601071213696 & 263.7604 & -49.4407 & 10.93 & 0.61 & B3Ie & 1/2 & -              \\
30 & CD-53  5736 & 5893945588395282304 & 223.1197 & -54.2952 & 10.87 & 2.53 & A0Ie & 1/2/3 & -          \\
31 & CD-54  5573 & 5896479309853592448 & 212.6621 & -55.0075 & 10.29 & 1.47 & A3I & 1/2 & -               \\
32 & CD-54  6746 & 5932016212933920384 & 246.1642 & -54.6357 & 9.45 & 1.11 & B8Iab & 1/2 & -              \\
33 & CD-55  5174 & 6063703586653222144 & 202.4626 & -56.1149 & 10.71 & 1.26 & B1Iae & 1/2/3 & -         \\
34 & CD-59  6142 & 5831295999979910656 & 246.2609 & -60.059 & 9.98 & 0.59 & A3Ie & 1/2/3 & -            \\
35 & Cl* NGC 6779    SAW V6 & 2039259886717168896 & 289.1491 & 30.1941 & 12.53 & 0.56 & kF5hF8 & 1 & -       \\
36 & EM* GGR   44 & 2005246464463628800 & 331.0513 & 53.0671 & 12.43 & 1.11 & B1I & 1/2 & R(1)                \\
37 & EM* StHA  161 & 2049984454412871296 & 290.4804 & 35.0486 & 11.31 & 0.27 & Fe & 1/2/3 & -           \\
38 & EM* VES  351 & 2168803045330976768 & 314.7316 & 49.5203 & 11.18 & 2.06 & F3Ie & 1/2/3 & -          \\
39 & HD  53300 & 3101342596792542464 & 106.0812 & -5.3054 & 7.94 & 0.98 & A2II & 1 & -                       \\
40 & HD  56126 & 3156171118495247360 & 109.0427 & 9.9967 & 8.1 & 0.08 & F0/5Ia & 1/2/3 & -              \\
41 & HD  93662 & 5351069693654349952 & 161.9101 & -57.4674 & 5.66 & 0.37 & K5 & 1/2 & B(1,2)                 \\    
42 & HD 101584 & 5343168568718268800 & 175.245 & -55.5738 & 6.92 & 0.82 & F0Iape & 1/2 & B(1,2,8)               \\    
43 & HD 105262 & 3920735495441657728 & 181.7951 & 12.9855 & 7.07 & 0.05 & B9 Ib & 1/2 & B(8)                 \\
44 & HD 107369 & 3469106382752903168 & 185.1873 & -32.5573 & 9.54 & 0.21 & A2II/III & 1/2 & -             \\
45 & HD 108015 & 6130448958959242240 & 186.2229 & -47.1521 & 7.86 & 0.31 & F3/5Ib/II & 1/2 & B(1,2)           \\    
46 & HD 116745 & 6083719439934104832 & 201.6097 & -47.2743 & 10.68 & 0.36 & F0Ibp & 1/2 & -               \\
47 & HD 133656 & 5903310335089068416 & 226.8643 & -48.2983 & 7.49 & 0.74 & A1/A2Ib/II & 1/2/3 & -       \\
48 & HD 144941 & 6042510190769087744 & 242.3523 & -27.2273 & 10.09 & 0.72 & B8 & 1 & -                       \\
49 & HD 148743 & 4351018375858237952 & 247.6251 & -7.5145 & 6.37 & 0.67 & A7Ib & 1/2 & -                  \\
50 & HD 157350 & 4122877783340594176 & 260.8558 & -17.971 & 8.56 & 0.1 & A2III/IV & 1 & -                    \\
51 & HD 161796 & 1367102319545324288 & 266.2311 & 50.0443 & 7.2 & 0.11 & F3Ib & 1/2/3 & -               \\
52 & HD 167402 & 4049624646596488576 & 274.0779 & -30.1249 & 8.94 & 0.73 & O9.5/B0Ib/II & 1 & R(1)               \\
53 & HD 172324 & 2096072103492979584 & 279.4949 & 37.4349 & 8.16 & 0.11 & A0Iabe & 1/2 & -                \\
54 & HD 172481 & 4072427555640528000 & 280.404 & -27.9503 & 8.84 & 0.65 & F2/3Ia & 1/2 & B(1,2)               \\    
55 & HD 177566 & 6715619076008049792 & 286.7829 & -41.7211 & 10.14 & 0.26 & B6Ib & 1/2 & R(1)                 \\
56 & HD 179821 & 4264026012336768000 & 288.4942 & 0.1255 & 7.55 & 1.25 & $G4_0-Ia$ & 1/2/3 & -          \\
57 & HD 186438 & 2049034819957965312 & 295.7205 & 37.6782 & 7.83 & 0.26 & F3Ib & 1/2 & -                  \\
58 & HD 187885 & 6871175064823382912 & 298.2196 & -17.0307 & 8.48 & 0.62 & F0Ie & 1/2/3 & -             \\
59 & HD 214539 & 6385794694664872320 & 340.1999 & -67.6886 & 7.2 & 0.09 & B8/9I & 2 & -                      \\
60 & HD 235858 & 2006425553228658816 & 337.2933 & 54.8517 & 8.2 & 0.48 & G5Ia & 1/2 & -                   \\
61 & HD 246299 & 3336558507975208448 & 85.2377 & 10.2403 & 10.27 & 0.58 & G2I & 1/2/3 & -               \\
62 & HD 306753 & 5335709477519159936 & 174.4288 & -60.8976 & 12.36 & 2.01 & A0 & 1/2 & -                  \\
63 & IRAS 01005+7910 & 565507868441719424 & 16.1896 & 79.4462 & 10.96 & 0.42 & B2Iab & 1/2/3 & -        \\
64 & IRAS 01259+6823 & 532078488712794624 & 22.3892 & 68.6547 & 11.81 & 2.74 & F5Ie & 2 & -                  \\
65 & IRAS 02528+4350 & 433515788197481984 & 44.0473 & 44.0478 & 10.75 & 0.34 & A0e & 1/2 & -              \\
66 & IRAS 07227-1320 & 3032030620730261376 & 111.2628 & -13.4389 & 11.6 & 0.66 & M1I & 2 & -                 \\
67 & IRAS 07582-4059 & 5534265613756612224 & 119.9905 & -41.1231 & 14.55 & 2.5 & - & 2/3 & R(2)              \\
68 & IRAS 08242-3828 & 5540178478053582592 & 126.5158 & -38.6465 & 12.03 & - & - & 1/2 & -                \\
69 & IRAS 08275-6206 & 5277809440015969792 & 127.1014 & -62.2724 & 10.8 & 0.69 & - & 1/2 & -              \\
70 & IRAS 08351-4634 & 5521628033275348480 & 129.19 & -46.7469 & 17.13 & 1.64 & - & 1/2 & R(3)               \\
71 & IRAS 09370-4826 & 5409357863031443840 & 144.728 & -48.6731 & 13.75 & 0.41 & - & 2 & R(3)                   \\
72 & IRAS 11387-6113 & 5335675087769798272 & 175.2863 & -61.5048 & 11.4 & 1.41 & A3Ie & 1/2/3 & -       \\
73 & IRAS 11531-6111 & 5335102207846402176 & 178.9084 & -61.4713 & 14.49 & 2.65 & - & 2 & -                  \\
74 & IRAS 12145-5834 & 6071416385848395008 & 184.3171 & -58.8582 & 15.09 & 2.55 & B8Ie & 2 & R(4)               \\
75 & IRAS 12360-5740 & 6060828565581083264 & 189.7213 & -57.9422 & 12.08 & 1.62 & - & 1/2 & -             \\
76 & IRAS 13110-6629 & 5857811238294426752 & 198.6128 & -66.7594 & 10.54 & - & - & 1/2 & -                \\
77 & IRAS 13356-6249 & 5865398796206273152 & 204.7763 & -63.079 & 15.8 & - & - & 1/2 & -                  \\
78 & IRAS 13421-6125 & 5865808020691983104 & 206.392 & -61.6677 & 16.18 & 5.47 & - & 2 & R(3)                   \\
79 & IRAS 14527-6204 & 5874676853324862720 & 224.1857 & -62.2815 & 11.08 & 1.74 & - & 2 & R(5)                  \\
80 & IRAS 15066-5532 & 5886573569080505216 & 227.6111 & -55.7367 & 14.57 & 5.55 & - & 1/2 & -             \\
81 & IRAS 16086-5255 & 5933063252888145920 & 243.1269 & -53.0528 & 13.15 & 3.28 & - & 1/2 & -             \\
82 & IRAS 16115-5044 & 5935061172876722176 & 243.8248 & -50.8721 & 14.83 & - & - & 2 & -                     \\
83 & IRAS 16476-1122 & 4334241408966611328 & 252.6012 & -11.466 & 11.09 & 1.7 & M1I & 2 & -                  \\
84 & IRAS 16494-3930 & 5969973999973524224 & 253.233 & -39.5818 & 16.73 & 1.83 & G2I & 1 & -                 \\
85 & IRAS 16594-4656 & 5963059480546004608 & 255.792 & -47.0077 & 14.6 & - & - & 1/2 & -                  \\
86 & IRAS 17208-3859 & 5972489407685976320 & 261.0812 & -39.0294 & 15.46 & 2.45 & - & 2 & -                  \\
87 & IRAS 17223-2659 & 4109553493474085504 & 261.361 & -27.0337 & 15.14 & 4.36 & - & 2 & -                   \\
88 & IRAS 17287-3443 & 5975119332093959552 & 263.0201 & -34.7591 & 12.92 & 4.53 & - & 2 & B(6)                  \\
89 & IRAS 17310-3432 & 4053542580087893376 & 263.585 & -34.5815 & 15.77 & 3.16 & - & 2 & R(1)                   \\
90 & IRAS 17332-2215 & 4117592469707529856 & 264.0712 & -22.2889 & 15.22 & 2.27 & - & 2 & -                  \\
91 & IRAS 17364-1238 & 4161796857143755264 & 264.8205 & -12.6749 & 12.8 & 1.76 & - & 2 & -                   \\
92 & IRAS 17433-1750 & 4120632688077368192 & 266.5659 & -17.8628 & 13.33 & 1.64 & M2I & 1/2 & R(3)           \\
93 & IRAS 17543-3102 & 4044070253255414656 & 269.39 & -31.051 & 14.84 & - & - & 2 & -                        \\
94 & IRAS 17579-3121 & 4043901443659288448 & 270.3057 & -31.3657 & 11.2 & - & - & 1/3 & -                 \\
95 & IRAS 17581-2926 & 4062288993280721792 & 270.314 & -29.4441 & 11 & 1.59 & - & 2 & R(5)                      \\
96 & IRAS 18084-1737 & 4095941436385621888 & 272.8678 & -17.611 & 15.71 & 1.07 & G3I & 1 & R(4)                 \\
97 & IRAS 18113-2503 & 4065347387968755328 & 273.6136 & -25.0501 & 15.07 & 3.01 & - & 1/2 & R(3)             \\
98 & IRAS 18158-3445 & 4044520262548753152 & 274.8057 & -34.7417 & 12.45 & 0.42 & F6 & 1 & B(7)                 \\
99 & IRAS 18435-0052 & 4260301176152524032 & 281.5326 & -0.8114 & 10.93 & 3.28 & B2II & 2 & -                \\
100 & IRAS 19075+0432 & 4293369057089082112 & 287.4997 & 4.619 & 14.44 & 4.79 & - & 1/2 & -               \\
101 & IRAS 19225+3013 & 2038872686817523072 & 291.1122 & 30.3241 & 12.24 & 0.67 & M2II & 2 & -               \\
102 & IRAS 19454+2920 & 2031794791233840128 & 296.8534 & 29.4697 & 15.29 & 4.43 & C-rich & 1/2 & -        \\
103 & IRAS 20094+3721 & 2060806470651334912 & 302.82 & 37.5145 & 10.63 & 1.24 & - & 1/2 & B(7)               \\
104 & IRAS 20174+3222 & 2054521833963867008 & 304.8659 & 32.5376 & 15.26 & - & - & 1/2 & -                \\
105 & IRAS 20244+3509 & 2056435602670418688 & 306.6049 & 35.3204 & 13.97 & 5.34 & - & 1/2 & -             \\
106 & IRAS 20259+4206 & 2068126263125039488 & 306.9261 & 42.2789 & 13.54 & 1.32 & - & 2 & -                  \\
107 & IRAS 20490+5934 & 2193902559325301760 & 312.5566 & 59.7642 & 10.37 & 0.63 & A3e & 1/2 & -           \\
108 & IRAS 21289+5815 & 2179471159976791168 & 322.5951 & 58.4811 & 14.46 & 1.35 & A2Ie & 1/2 & -          \\
109 & IRAS 21525+5643 & 2198987491374918528 & 328.5631 & 56.957 & 16.69 & 2.53 & - & 2 & -                   \\
110 & LB  3193 & 4715635535640762240 & 19.7214 & -61.9281 & 12.66 & 0.05 & - & 1/2 & -                    \\
111 & LS  IV -04    1 & 4365451214021224320 & 254.1155 & -4.7899 & 12 & 0.86 & B & 1/2 & -                \\
112 & LS  IV -15    3 & 4136944866387751552 & 260.7996 & -15.6209 & 11.76 & 1.21 & A0Ie & 1/2/3 & -     \\
113 & LS 4331 & 4124125282361429504 & 265.2502 & -16.3035 & 13.08 & 1.39 & B1Ibe & 1/2 & -                \\
114 & LS 5112 & 4099619470274753408 & 280.2026 & -17.0773 & 11.88 & 1.32 & B1IIIep & 1/2 & -              \\
115 & LS II +34 26 & 1869422453048750336 & 312.0693 & 34.4567 & 11.05 & 0.62 & B1.5Ia & 2 & R(4)                \\      %IRAS 20462+3416
116 & LSE  63 & 6736747708089687936 & 280.0917 & -31.9469 & 12.04 & 0.42 & B1Iabe & 1/2 & -               \\
117 & NGC  6254  1035 & 4365635279142583424 & 254.299 & -4.0666 & 11.43 & 0.84 & G0e & 1 & -                 \\
118 & OH 15.7 +0.8 & 4146237904302941440 & 274.1057 & -14.921 & 16.29 & 1.06 & - & 2 & R(3)                     \\      %IRAS 18135-1456
119 & OH 17.7 -2.0 & 4104128533107792512 & 277.6288 & -14.4793 & 13.65 & 1.55 & - & 2 & R(3)                    \\      %IRAS 18276-1431
120 & OH 345.05 -1.86 & 5965644393737729664 & 258.0907 & -42.4193 & 15.61 & 2.79 & - & 2 & R(3)                 \\
121 & PG 1704+222 & 4568163710366782848 & 256.6924 & 22.0978 & 12.69 & 0.23 & sdB3IHe8 & 2 & -               \\
122 & PHL  1580 & 6831062200578042624 & 322.6052 & -19.3762 & 12.18 & 0.1 & B2 & 1/2 & R(1)                  \\
123 & PN G038.7+01.5 & 4282499452616310912 & 284.0942 & 5.8833 & 12.5 & 2.91 & - & 2 & R(4)                     \\       %IRAS 18539+0549
124 & PN PM 1-243 & 4104509518206051968 & 278.7396 & -13.9802 & 14.43 & 1.44 & - & 2 & R(4)                     \\     %IRAS 18321-1401
125 & RAFGL 6945S & 4093773852321518976 & 272.8321 & -21.9166 & 13.99 & 1.7 & - & 2 & R(3)                      \\     %IRAS 18083-2155
126 & SS 441 & 4210278482327706496 & 294.073 & -3.8903 & 13.06 & - & - & 1 & -                               \\
127 & V* AD Aql & 4203801018864386944 & 284.7862 & -8.1706 & 11.28 & 0.78 & kF1hF5cnG5Ib & 1 & B(7)/V(1)             \\
128 & V* AU Vul & 1836195688380634368 & 304.5245 & 27.7343 & 9.98 & 1.57 & F3Ie & 1/2/3 & B(7)/V(1)             \\
129 & V* BZ Pyx & 5636099047820103424 & 137.0422 & -28.3196 & 10.91 & 0.35 & F6Ia & 1 & B(7)/V(1)                    \\
130 & V* CE Vir & 3658327596544582400 & 207.3213 & -1.9291 & 8.31 & 0.15 & G8III & 1 & B(1)                     \\
131 & V* EQ Cas & 1993916856117284864 & 358.222 & 55.0136 & 11.3 & - & - & 1 & -                             \\
132 & V* HP Lyr & 2101097215232231808 & 290.4128 & 39.9356 & 10.41 & 0.33 & A3Ia/Iab & 1 & B(7)/V(1)                 \\
133 & V* LN Hya & 3497154104039422848 & 194.1256 & -26.4603 & 6.62 & 0.14 & F3Ia & 1/2 & V(2)                \\
134 & V* LX And & 332909001084177920 & 34.9337 & 40.4562 & 15.71 & 0.07 & - & 1 & V(3)                          \\
135 & V* PS Gem & 3159640386918214528 & 105.9152 & 10.7703 & 7.24 & 0.08 & A0 & 1/2 & B(1,2,7)                  \\    
136 & V* RV Col & 2902505745786910080 & 83.9342 & -30.8265 & 8.41 & 0.15 & G5 & 1/2 & V(4)                   \\
137 & V* RX Cap & 6879691160336671744 & 303.7301 & -12.9429 & 11.37 & 0.28 & G0Iae & 1 & V(4)                    \\
138 & V* TT Oph & 4386330497453246080 & 252.3995 & 3.6317 & 9.85 & 0.23 & F5pe & 1 & V(4)                        \\
139 & V* TX Oph & 4392705672029913600 & 256.0004 & 4.9836 & 10.05 & 0.47 & F8Iae & 1 & -                     \\
140 & V* V1027 Cyg & 2030200671149815424 & 300.6141 & 30.0737 & 7.69 & 2.96 & G7Ia & 1/2 & -              \\
141 & V* V1333 Sco & 6023926760641310208 & 246.5849 & -34.2869 & 10.95 & 1.79 & F8 & 1 & B(7)                   \\
142 & V* V1401 Aql & 4190636669164572928 & 301.2726 & -11.5994 & 6.21 & 0.33 & F2II & 1/2 & -             \\
143 & V* V2053 Oph & 4470790101628029440 & 273.7058 & 5.2155 & 9.65 & 0.63 & C & 1 & B(1,7)                        \\ 
144 & V* V340 Ser & 4162959693758887424 & 262.6955 & -11.3689 & 9.3 & 1.31 & F2/3II & 2 & B(7)                  \\    %IRAS 17279-1119
145 & V* V360 Cyg & 1852749557493420032 & 317.6479 & 30.6724 & 11.08 & 0.4 & F8Ie & 1 & -                    \\
146 & V* V399 Cyg & 1869102495165230592 & 312.2852 & 33.6958 & 10.96 & 0.6 & G8 & 1 & V(4)                       \\
147 & V* V400 Sco & 4040579578519065728 & 267.8291 & -36.1824 & 12.7 & 1.7 & - & 1 & -                       \\
148 & V* V421 CMa & 5617989266685365120 & 109.0345 & -23.4504 & 10.49 & 0.95 & F5 & 1/2 & B(1,2,7)/V(1)              \\    
149 & V* V590 Aql & 4222177328438155776 & 304.2856 & -4.0519 & 11.84 & 0.47 & - & 1 & V(4)                      \\
150 & V* V652 Her & 4449366151908979072 & 252.0196 & 13.2618 & 10.51 & 0.1 & - & 1 & R(1)                       \\
151 & V* V709 Car & 5258718997589107712 & 154.8203 & -57.3239 & 9.3 & 1.41 & G8Ia-0 & 1 & B(1,7)/V(4)                    \\ 
152 & V* V760 Sgr & 4068898810519269248 & 267.5448 & -22.848 & 10.03 & 2.04 & G5 & 1 & -                     \\
153 & V* V802 Car & 5241806275407841664 & 165.518 & -62.1619 & 8.62 & 1.75 & F2III & 1/2 & B(1,2,7)/V(4)              \\    
154 & V* V811 Ara & 5914387846002855296 & 255.9235 & -61.5047 & 10.68 & 0.38 & - & 1 & -                     \\
155 & V* V825 Ara & 5921745812182394496 & 265.0453 & -53.7927 & 11.02 & 0.49 & - & 1 & -                     \\
156 & V* V956 Cen & 6066902993687172608 & 198.5344 & -54.6929 & 7.99 & 0.95 & F5Ia/ab & 1/2 & B(7)/V(4)            \\
157 & V* YY Ara & 5830750401670303872 & 250.3352 & -59.8752 & 9.47 & 0.58 & K0:-Me & 1 & -                   \\

\hline

\end{longtable}

\tablefoot{\tiny{Interstellar extinction values ({$A_{V}^{IS}$}) are taken from \citet{2019yCat.4038....0S}. Spectral types are taken from Simbad database. 
\textbf{Post-AGB origin:} (1): Simbad database. (2): Torun catalogue. (3): \citet{2006A&A...458..173S}. 
\textbf{Binary reference (B)}: (1): \citet{2019A&A...631A.108K}. (2): Torun catalogue. (3): \citet{2023A&A...671A..80G}. (4): \citet{2014A&A...563L..10V}. (5): \citet{2018A&A...620A..85O} (6) \citet{2016MNRAS.458.2565D}. (7): \citet{2022A&A...658A..36K}, (8): \citet{2022RNAAS...6..171P}. 
\textbf{Variability reference (V)}: (1): \citet{2022A&A...658A..36K}. (2): \citet{2022ApJ...927L..13K}. (3): \citet{2018A&A...617A..26D}. (4): Simbad database. 
\textbf{Removed reason (R)}: (1): very hot star ($T_{eff} \geq 24,000$ K). (2): incorrect source identification. (3): without \textit{Gaia} counterpart. (4): already in planetary nebula phase. (5): composite object, blended.}
}

%\newpage

% TABLE 1

\myfontsize

\begin{table}[h!]   
\caption{Astrometric and evolutionary parameters for the 69 post-AGB stars.} 
\label{table:postAGB}   

\myfontsize

\begin{tabular}  {l c c c c c c c c c c c c c}    
\hline\hline                   
 Num & Distance & Low Dist. & High Dist. & $A_V$ & $T_{eff}$ & Flag (T) & $Log[L]$ & $Log[L]_{min}$ & $Log[L]_{max}$ & $Mass$ & $Age_{evo}$ &  SED & Flag	\\  
  & (pc) & (pc) & (pc) & & (K) & & & &  & ($M_{\odot}$) & (kyr) & &\\ 
\hline

2 & 1307 & 1199 & 1436 & 0.50 & 6500 & T3(1) & 4.11 & 4.02 & 4.20 & 3.14 & $<0.01$ & disc & -				\\
7 & 1909 & 1870 & 1947 & 3.85 & 4250 & T3(2) & 4.34 & 4.32 & 4.37 & 2.79 & 1.22 & stellar & H            \\
14 & 6639 & 5922 & 7516 & 1.75 & 16000 & T3(3) & 3.75 & 3.63 & 3.88 & 1.33 & 3.62 & stellar & -          \\
15 & 2271 & 2160 & 2400 & 4.75 & 3750 & T1 & 3.61 & 3.56 & 3.66 & 1.12 & $<0.01$ & stellar & H          \\
18 & 3467 & 3159 & 3934 & 0.20 & 20000 & T3(4) & 3.55 & 3.44 & 3.67 & 0.93 & 18.14 & shell & H           \\
19 & 2839 & 2711 & 2962 & 1.60 & 6500 & T3(5) & 3.62 & 3.58 & 3.66 & 1.12 & 5.61 & shell & -             \\
21 & 4734 & 4150 & 5414 & 0.30 & 21000 & T3(6) & 3.61 & 3.47 & 3.75 & 0.95 & 16.54 & shell & H           \\
24 & 9965 & 9131 & 11304 & 0.90 & 22000 & T3(7) & 4.30 & 4.19 & 4.42 & 3.81 & 1.10 & stellar & (*)         \\
25 & 5862 & 5008 & 7121 & 0.73 & 23000 & T3(8) & 4.40 & 4.21 & 4.58 & 3.00 & 9.15 & stellar & H          \\
26 & 4982 & 4299 & 6049 & 1.00 & 18000 & T3(9) & 4.24 & 4.06 & 4.43 & 3.63 & 0.99 & stellar & -          \\
27 & 4718 & 4214 & 5732 & 3.30 & 6000 & T2 & 4.12 & 3.95 & 4.29 & 3.16 & $<0.01$ & disc & -             \\
29 & 3858 & 3644 & 4143 & 2.15 & 20000 & T3(6) & 4.43 & 4.37 & 4.49& 4.00 & 0.69 & shell &-            \\
30 & 3421 & 3226 & 3635 & 3.14 & 9250 & T2 & 3.87 & 3.81 & 3.93 & 1.98 & 3.36 & shell & -             \\
31 & 4340 & 4100 & 4599 & 1.97 & 9000 & T2 & 3.77 & 3.72 & 3.83 & 1.42 & 3.94 & shell & -             \\
33 & 2822 & 2677 & 2982 & 1.78 & 20000 & T3(6) & 4.02 & 3.97 & 4.08 & 2.84 & 2.14 & shell & -            \\
34 & 5021 & 4607 & 5683 & 1.00 & 8500 & T2 & 3.76 & 3.65 & 3.88 & 1.38 & 3.82 & shell & -             \\
35 & 10125 & 9128 & 11379 & 2.25 & 6750 & T2 & 3.67 & 3.56 & 3.78 & 0.97 & 9.06 & stellar & H         \\
37 & 10312 & 8988 & 12281 & $<0.01$ & 11750 & T2 & 3.45 & 3.28 & 3.63 & $<0.9$ & 13.83 & shell & H         \\
39 & 2916 & 2712 & 3196 & 0.80 & 7250 & T3(10) & 3.92 & 3.84 & 4.01 & 2.27 & 1.49 & stellar & (*)           \\
40 & 2099 & 1991 & 2209 & 2.00 & 7250 & T3(11) & 3.94 & 3.89 & 3.99 & 2.40 & 1.63 & shell & H             \\
41 & 1104 & 1075 & 1139 & 1.00 & 4250 & T3(12) & 4.16 & 4.13 & 4.19 & 3.22 & $<0.01$ & disc & -             \\
42 & 1788 & 1722 & 1845 & 0.82 & 7250 & T3(13) & 3.95 & 3.92 & 3.98 & 2.42 & 1.66 & disc & -              \\
43 & 1567 & 1503 & 1632 & 0.25 & 8250 & T3(14) & 3.54 & 3.50 & 3.58 & 0.93 & 9.14 & stellar & H           \\
45 & 5130 & 4530 & 5923 & 0.70 & 7000 & T3(13) & 4.40 & 4.27 & 4.53 & 3.00 & 9.14 & disc & H              \\
47 & 1708 & 1647 & 1782 & 1.00 & 8250 & T3(11) & 3.74 & 3.71 & 3.77 & 1.32 & 3.57 & shell & -             \\
49 & 3101 & 2895 & 3388 & 0.60 & 6750 & T3(11) & 4.51 & 4.44 & 4.58 & 3.00 & 9.14 & stellar & H           \\
51 & 1921 & 1830 & 2016 & 0.75 & 6000 & T3(11) & 3.88 & 3.83 & 3.93 & 2.07 & 0.28 & shell & H             \\
53 & 1807 & 1726 & 1891 & 0.40 & 9750 & T3(15) & 3.38 & 3.34 & 3.42 & $<1.0$ & 9.67 & stellar & -           \\
54 & 6999 & 5822 & 8254 & 1.40 & 7000 & T3(16) & 4.56 & 4.37 & 4.74 & 3.00 & 9.14 & disc & H              \\
58 & 2310 & 2165 & 2481 & 1.80 & 8000 & T3(11) & 3.89 & 3.83 & 3.95 & 2.12 & 1.34 & shell & -             \\
59 & 1390 & 1344 & 1439 & 0.35 & 10000 & T3(17) & 3.57 & 3.54 & 3.60 & 1.07 & 5.54 & stellar & H (*)          \\
60 & 1410 & 1356 & 1465 & 2.80 & 5250 & T3(11) & 3.94 & 3.90 & 3.98 & 2.35 & 0.58 & shell & -             \\
62 & 5416 & 5040 & 5792 & 2.19 & 14000 & T3(18) & 3.51 & 3.44 & 3.58 & $<1.0$ & 10.68 & shell & -           \\
63 & 3685 & 3461 & 3942 & 1.80 & 22000 & T3(19) & 4.15 & 4.09 & 4.22 & 3.32 & 0.79 & shell & -            \\
64 & 4817 & 4587 & 5042 & 3.00 & 5500 & T3(11) & 3.47 & 3.43 & 3.51 & $<1.0$ & 8.54 & shell & -             \\
72 & 5022 & 4555 & 5658 & 3.35 & 9000 & T3(18) & 3.99 & 3.88 & 4.09 & 2.62 & 1.88 & shell & -             \\
75 & 9082 & 8261 & 10230 & 2.70 & 7500 & T3(11) & 3.88 & 3.78 & 3.97 & 2.08 & 1.29 & shell & -            \\
98 & 6605 & 5703 & 7346 & 2.25 & 6750 & T2 & 3.38 & 3.25 & 3.51 & $<1.0$ & 9.14 & disc & -              \\
99 & 2082 & 2026 & 2148 & 3.75 & 14000 & T3(20) & 3.91 & 3.88 & 3.94 & 2.23 & 1.45 & shell & -            \\
102 & 6830 & 5887 & 8163 & 6.75 & 9750 & T3(18) & 3.69 & 3.51 & 3.87 & 1.20 & 8.33 & shell & -            \\
111 & 10863 & 9455 & 12375 & 1.40 & 15000 & T3(21) & 4.23 & 4.09 & 4.38 & 2.58 & 1.83 & stellar & H       \\
112 & 7934 & 6777 & 9233 & 1.75 & 19000 & T3(6) & 4.41 & 4.25 & 4.57 & 3.00 & 1.84 & shell & H           \\
113 & 6437 & 5646 & 7505 & 1.60 & 16000 & T3(22) & 3.44 & 3.29 & 3.60 & $<1.0$ & 11.32 & shell & H          \\
114 & 5354 & 4801 & 6044 & 2.10 & 19000 & T3(23) & 4.01 & 3.89 & 4.13 & 2.77 & 2.07 & shell & -           \\
116 & 7318 & 6417 & 8628 & 1.05 & 22000 & T3(6) & 4.12 & 3.96 & 4.29 & 2.26 & 1.55 & shell & H           \\
117 & 5974 & 5146 & 7444 & 1.00 & 5750 & T3(24) & 3.22 & 3.00 & 3.43 & 0.90 & 7.96 & stellar & H          \\
121 & 7513 & 6290 & 9363 & 0.30 & 17000 & T3(25) & 3.34 & 3.11 & 3.58 & $<0.9$ & 19.81 & stellar & H        \\
127 & 6147 & 5365 & 6967 & 1.20 & 6250 & T3(13) & 3.46 & 3.32 & 3.61 & $<1.0$ & 8.96 & disc & -             \\
128 & 2317 & 2237 & 2410 & 3.50 & 5750 & T3(26) & 3.73 & 3.69 & 3.78 & 1.24 & 8.85 & disc & -             \\
129 & 4556 & 4246 & 4929 & 1.35 & 6750 & T3(27) & 3.34 & 3.26 & 3.42 & $<1.0$ & 9.14 & disc & -             \\
132 & 10720 & 9413 & 12080 & 1.75 & 8500 & T2 & 4.48 & 4.35 & 4.61 & 3.00 & 9.15 & disc & H           \\
133 & 1684 & 1619 & 1769 & 1.00 & 6250 & T3(11) & 4.02 & 3.97 & 4.06 & 2.71 & 1.92 & disc & -             \\
135 & 1813 & 1707 & 1923 & 2.40 & 6000 & T3(28) & 4.49 & 4.43 & 4.55 & 4.00 & 4.05 & disc & -             \\
136 & 1850 & 1806 & 1893 & 2.25 & 5500 & T2 & 3.80 & 3.78 & 3.83 & 1.62 & 1.53 & shell & -            \\
137 & 7253 & 6109 & 8872 & 0.70 & 6000 & T2 & 3.40 & 3.17 & 3.64 & $<0.9$ & 8.29 & stellar & H          \\
138 & 2369 & 2250 & 2507 & 2.25 & 6500 & T2 & 3.46 & 3.40 & 3.53 & $<1.0$ & 8.96 & stellar & -          \\
139 & 4682 & 4285 & 5128 & 1.75 & 6250 & T2 & 3.86 & 3.77 & 3.95 & 1.18 & 7.64 & shell & H            \\
141 & 3948 & 3726 & 4264 & 2.00 & 6250 & T3(29) & 3.37 & 3.31 & 3.43 & $<1.0$ & 8.96 & disc & -             \\
142 & 727 & 714 & 743 & 1.00 & 6750 & T3(11) & 3.47 & 3.45 & 3.49 & $<1.0$ & 9.14 & shell & -               \\
143 & 4188 & 3841 & 4631 & 1.20 & 5000 & T3(27) & 3.67 & 3.58 & 3.77 & 1.18 & 7.13 & disc & -             \\
144 & 4269 & 3980 & 4580 & 2.00 & 7250 & T2 & 4.10 & 4.03 & 4.16 & 3.13 & 0.67 & disc & -             \\
145 & 4795 & 4442 & 5245 & 1.50 & 6250 & T2 & 3.41 & 3.32 & 3.51 & $<1.0$ & 8.96 & stellar & -          \\
146 & 4842 & 4502 & 5201 & 2.25 & 5500 & T2 & 3.58 & 3.51 & 3.65 & 1.08 & 4.78 & stellar & -          \\
147 & 8617 & 7584 & 9718 & 1.70 & 4500 & T2 & 3.34 & 3.21 & 3.47 & $<1.0$ & 6.83 & stellar & -          \\
148 & 5122 & 4745 & 5526 & 10.25 & 7000 & T3(30) & 4.13 & 4.06 & 4.21 & 3.22 & $<0.01$ & disc & -           \\
152 & 2116 & 1995 & 2245 & 2.80 & 5500 & T3(31) & 3.40 & 3.34 & 3.47 & $<1.0$ & 8.54 & stellar & -          \\
153 & 4404 & 4113 & 4693 & 2.00 & 7500 & T3(27) & 4.42 & 4.35 & 4.48 & 4.00 & 9.14 & disc & -             \\
156 & 1535 & 1468 & 1606 & 1.80 & 6750 & T2 & 3.64 & 3.60 & 3.68 & 1.14 & 6.95 & shell & -            \\
157 & 2348 & 2251 & 2450 & 2.80 & 4250 & T3(32) & 3.41 & 3.37 & 3.46 & $<1.0$ & 5.58 & stellar & -          \\

\hline

\end{tabular}

\tablefoot{\tiny{(*): Not confirmed previously as post-AGB star in the bibliography.
(H): Suspected Halo star.}}

\tablebib{
\tiny{Temperatures are: (\textbf{T1}) obtained from SED fitting with $A_{V}^{IS}$ from \citet{2019yCat.4038....0S}, (\textbf{T2}): derived from Simbad spectral types, (\textbf{T3}): obtained from the literature (mainly for spectroscopic measurements): (1): \citet{2014AJ....147..137L}, (2): \citet{2020AJ....160...83S}, (3): \citet{1998A&A...340..476J}, (4): \citet{1994A&A...292..239N}, (5): \citet{2019ApJ...879...69T}, (6): \citet{2012A&A...543A..11M}, (7): \citet{1998A&A...334..987V}, (8): \citet{2020MNRAS.494.2117H}, (9): \citet{1993A&A...279..188J}, (10): \citet{2012MNRAS.419.1254R}, (11): \citet{2022ApJ...927L..13K}, (12): \citet{2005A&A...435..161D}, (13): \citet{2023A&A...674A.151C}, (14): \citet{2019A&A...627A.138A}, (15): \citet{2018AstBu..73...52K}, (16): \citet{2001A&A...365..465R}, (17): \citet{1984ApJ...278..208K}, (18): \citet{2017MNRAS.470.1593R}, (19): \citet{2014AstBu..69..279K}, (20): \citet{2000AAS...197.6005P}, (21): \citet{2018AstBu..73..211S}, (22): \citet{2000A&AS..145..269P}, (23): \citet{2020MNRAS.491.4829I}, (24): \citet{1958ApJ...127..583W}, (25): \citet{2013A&A...551A..31D}, (26): \citet{2022A&A...658A..36K}, (27): \citet{2011A&A...533A..99G}, (28): \citet{1991A&A...251..495W}, (29): \citet{2005A&A...429..297M}, (30): \citet{1997A&A...319..561V}, (31): \citet{2021MNRAS.506..150B}, (32): \citet{1976ApJS...30..491H}. }
}

\end{table}

%\newpage

% TABLE 2

%\begin{center}

\small

\begin{table}[h!]   
\caption{Astrometric and evolutionary parameters for the 29 unconfirmed post-AGB candidates.} 
\label{table:postRGB}      

\myfontsize

\begin{tabular}  {l c c c c c c c c c c c}    
\hline\hline                   
Num & Distance & Low Dist. & High Dist. & $A_V$ & T(eff) & Flag (T)  & $Log[L]$ & $Log[L]_{min}$ & $Log[L]_{max}$ & SED & Flag  \\   
&   (pc) & (pc) & (pc) & & (K)  &  & & &  & & \\  
\hline    

6 & 4007 & 3509 & 4603 & 0.09 & 3500 & T1 & 3.19 & 3.05 & 3.33 & stellar         & H \\
8 & 7877 & 7285 & 8653 & $<0.01$ & 7250 & T2 & 2.92 & 2.83 & 3.01 & stellar       & H \\
16 & 7633 & 6920 & 8251 & 1.40 & 11000 & T3(1) & 3.12 & 3.03 & 3.20 & stellar     & H \\
23 & 2569 & 2382 & 2815 & 0.20 & 11750 & T2 & 2.46 & 2.38 & 2.54 & stellar     & H \\
32 & 2138 & 2058 & 2229 & 0.70 & 9500 & T3(2) & 3.16 & 3.12 & 3.19 & stellar      & H \\
44 & 2568 & 2429 & 2706 & 0.30 & 7500 & T3(3) & 2.99 & 2.94 & 3.04 & stellar      & H \\
46 & 4893 & 4514 & 5338 & 0.60 & 6750 & T3(4) & 3.19 & 3.11 & 3.27 & stellar     & H \\
48 & 1446 & 1391 & 1510 & 0.85 & 22000 & T3(5) & 2.56 & 2.52 & 2.60 & stellar     & - \\
57 & 945 & 921 & 964 & 0.40 & 6500 & T3(6) & 2.84 & 2.82 & 2.86 & disc            & - \\
61 & 937 & 926 & 947 & 1.00 & 5750 & T3(7) & 2.06 & 2.05 & 2.07 & shell           & - \\
66 & 1982 & 1915 & 2056 & 0.66 & 3750 & T3(8) & 2.37 & 2.33 & 2.40 & shell        & - \\
69 & 1952 & 1895 & 2002 & 0.69 & 3750 & T1 & 2.63 & 2.60 & 2.65 & shell        & - \\
73 & 5266 & 4836 & 5721 & 4.00 & 1000 & T3(8) & 3.07 & 2.99 & 3.15 & shell       & - \\
80 & 3226 & 3025 & 3469 & 5.55 & 6000 & T1 & 2.76 & 2.70 & 2.83 & shell       & - \\
81 & 2504 & 2408 & 2629 & 3.28 & 6750 & T1 & 2.45 & 2.41 & 2.49 & shell       & - \\
83 & 1978 & 1856 & 2105 & 2.50 & 4000 & T3(9) & 3.01 & 2.95 & 3.06 & disc        & - \\
86 & 4277 & 3572 & 5089 & 5.80 & 8750 & T3(8) & 3.00 & 2.81 & 3.19 & shell       & - \\
87 & 5204 & 4054 & 6718 & 2.00 & 3500 & T3(8) & 1.96 & 1.64 & 2.27 & shell       & - \\
88 & 2533 & 2306 & 2812 & 4.53 & 3750 & T1 & 2.94 & 2.84 & 3.04 & shell       & - \\
90 & 4882 & 4170 & 5705 & 3.00 & 4250 & T3(8) & 2.27 & 2.10 & 2.43 & shell       & - \\
91 & 5991 & 5520 & 6694 & 1.76 & 8250 & T1 & 3.01 & 2.91 & 3.11 & shell       & - \\
100 & 4898 & 4294 & 5685 & 4.79 & 5750 & T1 & 3.01 & 2.87 & 3.16 & disc       & - \\
101 & 6087 & 5536 & 6736 & 0.67 & 3500 & T3(8) & 3.15 & 3.06 & 3.25 & shell       & - \\
105 & 1694 & 1631 & 1768 & 5.34 & 4250 & T1 & 2.50 & 2.46 & 2.53 & disc        & - \\
110 & 5710 & 4998 & 6635 & $<0.01$ & 14000 & T3(10) & 2.79 & 2.64 & 2.94 & stellar    & H \\
130 & 1365 & 1313 & 1420 & 0.90 & 4750 & T2 & 3.21 & 3.16 & 3.27 & shell       & - \\
149 & 6299 & 5704 & 7065 & 0.47 & 4000 & T3(9) & 3.06 & 2.95 & 3.16 & stellar     & H \\
154 & 4714 & 4378 & 5080 & 0.38 & 4000 & T1 & 3.29 & 3.22 & 3.36 & stellar     & - \\
155 & 2123 & 1943 & 2310 & 0.49 & 3500 & T1 & 2.88 & 2.80 & 2.96 & stellar     & - \\

\hline

\end{tabular}

\tablefoot{(H): Suspected Halo star.}

\tablebib{
\tiny {Temperatures are: (\textbf{T1}) obtained from SED fitting with $A_V$ from \citet{2019yCat.4038....0S}, (\textbf{T2}): derived from Simbad spectral types, (\textbf{T3}): obtained from the literature (mainly for spectroscopic measurements): (1): \citet{2004A&A...419.1123M}, (2): \citet{1998A&A...334..987V}, (3): \citet{2022ApJ...927L..13K}, (4): \citet{1992MNRAS.254..343G}, (5): \citet{1973A&A....25..261H}, (6): \citet{2022A&A...658A..36K}, (7): 
\citet{2019ApJ...879...69T}, (8): \citet{2006A&A...458..173S}, (9): \citet{2020AJ....160...83S}, (10): \citet{1992A&A...260..261Q}.}
}

\end{table}

%\newpage

% TABLE 3

\begin{table}[h!]   
\caption{Astrometric and evolutionary parameters for the 14 YSO candidate stars.} 
\label{table:YSO}      

\myfontsize

\begin{tabular}  {l c c c c c c c c c c c}    
\hline\hline                   
Num & Distance & Low Dist. & High Dist. &  $A_V$ & T(eff) & Flag (T)  & $Log[L]$ & $Log[L]_{min}$ & $Log[L]_{max}$ & SED & Flag  \\   
&   (pc) & (pc) & (pc) & & (K)  &  & & &  & & \\  
\hline

1 & 1139 & 1109 & 1172 & 1.45 & 9250 & T3(1) & 4.28 & 4.25 & 4.30 & stellar & MC \\
11 & 3379 & 3021 & 3936 & 0.79 & 5000 & T3(2) & 0.69 & 0.55 & 0.82 & stellar & -     \\
17 & 307 & 305 & 308 & 0.07 & 6250 & T2 & 1.10 & 1.09 & 1.10 & stellar & -        \\
38 & 623 & 619 & 628 & 2.75 & 7500 & T3(3) & 1.97 & 1.96 & 1.98 & disc & MC           \\
50 & 219 & 217 & 220 & 0.50 & 8750 & T2 & 1.40 & 1.39 & 1.40 & stellar & -        \\
65 & 390 & 387 & 393 & 0.93 & 9500 & T2 & 1.22 & 1.21 & 1.22 & uncertain & -      \\
%78 & 3645 & 3168 & 4215 & 5.47 & 4500 & T1 & 2.23 & 2.09 & 2.38 & shell & MC       \\
84 & 2915 & 2503 & 3473 & 2.20 & 5000 & T3(3) & 0.87 & 0.70 & 1.05 & shell & -       \\
103 & 1787 & 1746 & 1825 & 2.25 & 5500 & T3(4) & 2.88 & 2.85 & 2.90 & disc & MC       \\
106 & 1060 & 1046 & 1072 & 1.32 & 6750 & T3(3) & 0.97 & 0.95 & 0.98 & shell & -      \\
107 & 482 & 478 & 485 & 0.63 & 8500 & T3(2) & 1.38 & 1.37 & 1.39 & disc & -          \\
108 & 964 & 948 & 981 & 2.05 & 6750 & T3(3) & 0.90 & 0.89 & 0.92 & disc & MC          \\
109 & 2186 & 1944 & 2596 & 2.53 & 6000 & T1 & 0.72 & 0.56 & 0.87 & shell & -      \\
134 & 508 & 497 & 520 & 0.07 & 5500 & T1 & -0.94 & -1.01 & -0.87 & disc & -       \\
151 & 4006 & 3630 & 4403 & 6.75 & 5000 & T2 & 5.32 & 5.23 & 5.42 & disc & MC       \\

\hline

\end{tabular}

\tablefoot{\tiny{(MC): YSO located in a molecular cloud.}
%(**): Possible binary star. 
}

\tablebib{
\tiny{Temperatures are: (\textbf{T1}) obtained from SED fitting with $A_V$ from \citet{2019yCat.4038....0S}, (\textbf{T2}): derived from Simbad spectral types, (\textbf{T3}): obtained from the literature (mainly for spectroscopic measurements): (1): \citet{2012A&A...543A..80F}, (2): \citet{2006A&A...458..173S}, (3): \citet{2017MNRAS.470.1593R}, (4): \citet{2022A&A...658A..36K}.}
}

\end{table}

%\newpage

% TABLE 4

\begin{table}[h!]   
\caption{Astrometric and evolutionary parameters for the 6 stars classified as Supergiant or Horizontal Branch stars.}  
\label{table:supergiant}      

\small

\begin{tabular}  {l c c c c c c c c c c c}    
\hline\hline                   
Num & Distance & Low Dist. & High Dist. &  $A_V$ & T(eff) & Flag (T)  & $Log[L]$ & $Log[L]_{min}$ & $Log[L]_{max}$ & SED & Flag  \\   
&   (pc) & (pc) & (pc) & & (K)  &  & & &  & & \\  
\hline
4 & 3506 & 2918 & 4516 & 1.64 & 7250 & T2 & 1.06 & 0.79 & 1.32 & stellar & HB/H      \\
5 & 4503 & 3936 & 5076 & 0.41 & 9750 & T1 & 1.41 & 1.28 & 1.54 & stellar & HB/H          \\
20 & 3661 & 3424 & 3917 & 3.75 & 16000 & T2 & 4.69 & 4.62 & 4.75 & disc & S       \\
22 & 3604 & 3342 & 3931 & 0.06 & 10750 & T1 & 1.78 & 1.7 & 1.86 & stellar & HB/H     \\
56 & 4432 & 4078 & 4782 & 4.00 & 7500 & T3(1) & 5.47 & 5.4 & 5.55 & shell & S           \\
140 & 3723 & 3486 & 3988 & 3.8 & 5000 & T2 & 5.13 & 5.06 & 5.19 & disc & S        \\

\hline

\end{tabular}

\tablefoot{
\tiny{(HB): Horizontal Branch star. (S): Supergiant star. (H): Halo star}
}

\tablebib{
\tiny{Temperatures are: (\textbf{T1}) obtained from SED fitting with $A_V$ from \citet{2019yCat.4038....0S}, (\textbf{T2}): derived from Simbad spectral types, (\textbf{T3}): obtained from the literature (mainly for spectroscopic measurements): (1): \citet{2019A&A...627A.138A}.}
}

\end{table}

\newpage

\end{appendix}

\end{document}